\documentclass{article}

\usepackage[english]{babel}

\usepackage[letterpaper,top=2cm,bottom=2cm,left=3cm,right=3cm,marginparwidth=1.75cm]{geometry}

\usepackage{amsmath}
\usepackage{graphicx}
\usepackage[colorlinks=true, allcolors=blue]{hyperref}
\usepackage{graphicx}
\usepackage{float}
\usepackage{amsfonts} 
\usepackage[straightvoltages]{circuitikz}
\usepackage{tikz}
\usepackage[acronym,shortcuts]{glossaries}
\usepackage{xcolor}
\usetikzlibrary{shapes,arrows,fit,shapes.multipart,shapes.misc,fit}
\ctikzset{bipoles/resistor/height=0.25}
\ctikzset{bipoles/resistor/width=0.5}
\tikzstyle{block} = [draw, rectangle, 
    minimum height=3em, minimum width=6em]
\tikzstyle{sum} = [draw, circle, node distance=1cm]
\tikzstyle{input} = [coordinate]
\tikzstyle{output} = [coordinate]
\tikzstyle{pinstyle} = [pin edge={to-,thin,black}]
\usetikzlibrary{arrows.meta}

\newcommand{\vc}[1]{\mathbf{#1}}
\newcommand{\bs}[1]{\boldsymbol{#1}}
\newcommand{\pfrac}[2]{\frac{\partial #1}{\partial #2}}

\newcommand{\pluseq}{\mathrel{+}=}

\newcommand{\od}[1]{\mathcal{#1}}

\newcommand{\param}{\boldsymbol{\theta}}
\newcommand{\paramm}{\param_\text{MAP}}
\newcommand{\paramvmr}{\param_\text{VMR}}
\newcommand{\zparam}{\boldsymbol{\alpha}}
\newcommand{\zparamb}{\zparam_\text{b}}
\newcommand{\zparamj}{\zparam_\text{j}}
\newcommand{\zparamo}{\hat{\zparam}}
\newcommand{\zparamob}{\zparamo_\text{b}}
\newcommand{\zparamoj}{\zparamo_\text{j}}
\newcommand{\prob}{\mathfrak{p}}
\newcommand{\probz}{\prob_\text{0D}}
\newcommand{\probt}{\prob_\text{3D}}
\newcommand{\model}{\mathfrak{M}}
\newcommand{\modelz}{\model_\text{0D}}
\newcommand{\modelt}{\model_\text{3D}}

\newcommand{\yobs}{\boldsymbol{y}_{\text{obs}}}
\newcommand{\snr}{\text{SNR}}
\newcommand{\posterior}{\prob(\param|\yobs)}

\newacronym{CFD}{CFD}{Computational Fluid Dynamics}
\newacronym{SMC}{SMC}{Sequential Monte Carlo}
\newacronym{MAP}{MAP}{maximum a posteriori}
\newacronym{LPN}{LPN}{lumped-parameter network}
\newacronym{ROM}{ROM}{Reduced-order model}
\newacronym{BC}{BC}{boundary condition}
\newacronym{LM}{LM}{Levenberg-Marquardt}
\newacronym{uq}{UQ}{uncertainty quantification}
\newacronym{0D}{0D}{zero-dimensional}
\newacronym{1D}{1D}{one-dimensional}
\newacronym{3D}{3D}{three-dimensional}
\newacronym{mri}{MRI}{magnetic resonance imaging}

\newcommand{\reva}[1]{\textcolor{black}{#1}}
\newcommand{\revb}[1]{\textcolor{black}{#1}}

\definecolor{clslow}{HTML}{B1040E}
\definecolor{clfast}{HTML}{008566}
\definecolor{cloptimized}{HTML}{286dc0}
\definecolor{clgeometric}{HTML}{978d85}

\title{Bayesian Windkessel calibration using optimized 0D surrogate models}
\author{Jakob Richter,$^{1,2}$ Jonas Nitzler,$^{2,3}$ Luca Pegolotti,$^{1,4}$ Karthik Menon,$^{1,4}$ Jonas Biehler,$^2$ \\Wolfgang A. Wall,$^2$ Daniele E. Schiavazzi,$^5$ Alison L. Marsden,$^{1,4}$ Martin R. Pfaller$^{1,4}$}
\date{}

\begin{document}
\maketitle

\subsubsection*{Affiliations}
\begin{enumerate}
\small
\item Department of Pediatrics, Stanford University
\item Institute for Computational Mechanics, Technical University of Munich
\item Professorship for Data-driven Materials Modeling, Technical University of Munich
\item Institute for Computational and Mathematical Engineering, Stanford University
\item Department of Applied and Computational Mathematics and Statistics, University of Notre Dame
\end{enumerate}

\subsubsection*{Abstract}
Bayesian \gls*{BC} calibration approaches from clinical measurements have successfully quantified inherent uncertainties in cardiovascular fluid dynamics simulations. However, estimating the posterior distribution for all \gls*{BC} parameters in \gls*{3D} simulations has been unattainable due to infeasible computational demand. We propose an efficient method to identify Windkessel parameter posteriors: We only evaluate the \gls*{3D} model once for an initial choice of \glspl*{BC} and use the result to create a highly accurate \gls*{0D} surrogate. We then perform \gls*{SMC} using the optimized \gls*{0D} model to derive the high-dimensional Windkessel \gls*{BC} posterior distribution. Optimizing 0D models to match \gls*{3D} data a priori lowered their median approximation error by nearly one order of magnitude in 72 publicly available vascular models. The optimized \gls*{0D} models generalized well to a wide range of \glspl*{BC}. Using \gls*{SMC}, we evaluated the high-dimensional Windkessel parameter posterior for different measured signal-to-noise ratios in a vascular model, which we validated against a \gls*{3D} posterior. The minimal computational demand of our method using a single \gls*{3D} simulation, combined with the open-source nature of all software and data used in this work, will increase access and efficiency of Bayesian Windkessel calibration in cardiovascular fluid dynamics simulations.

\section{Introduction\label{sec_intro}}
\begin{figure}
\centering
\begin{minipage}[c]{3.6cm}
\centering
\gls*{3D}\\model $\modelt(\param)$\\[.5cm]
\includegraphics[width=\linewidth]{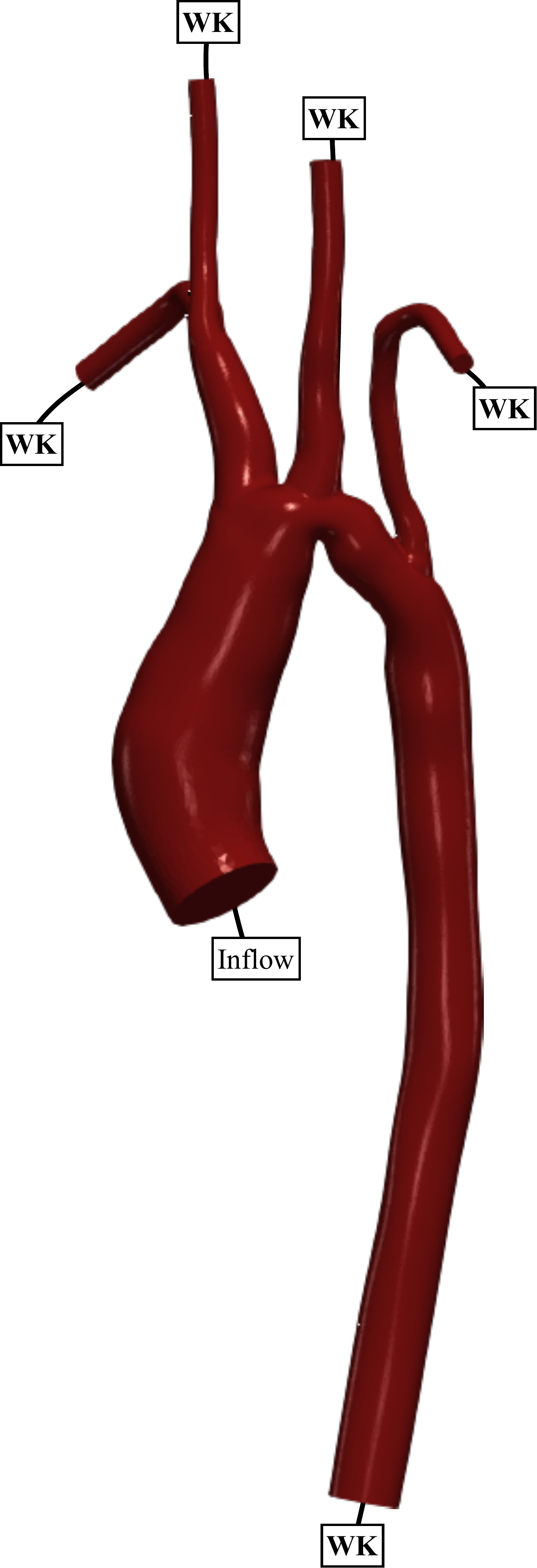}
\end{minipage}
\hspace{1cm}
\begin{minipage}[c]{3.6cm}
\centering
Geometric \gls*{0D} model $\modelz(\param,\zparamb)$\\[.5cm]
\includegraphics[width=\linewidth]{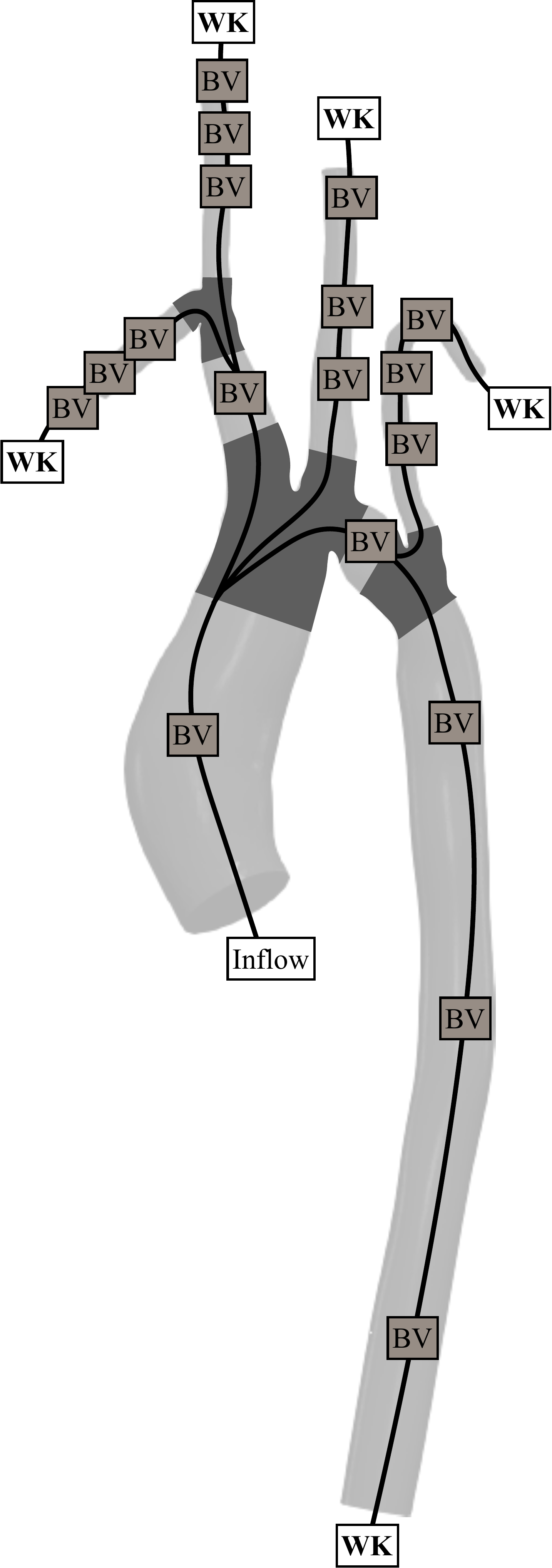}
\end{minipage}
\hspace{1cm}
\begin{minipage}[c]{3.6cm}
\centering
Optimized \gls*{0D} model $\modelz(\param,\zparamo)$\\[.5cm]
\includegraphics[width=\linewidth]{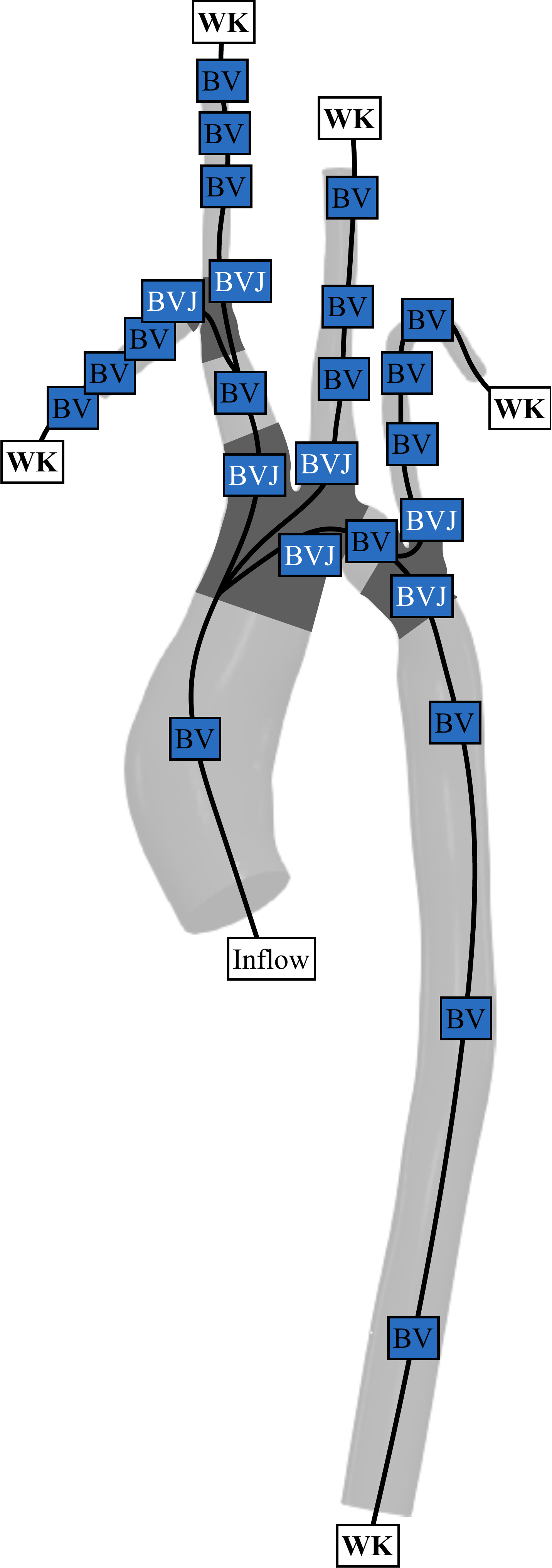}
\end{minipage}
\caption{\setlength{\fboxsep}{1pt} Illustration of different blood flow models used in this work. Inflow (\fbox{Inflow}) and \texttt{Windkessel} boundary conditions (\fbox{WK}) with parameters $\param$ are identical in all models. The geometry model is separated into branch (light) and junction (dark) regions. The \textit{geometric} 0D model (middle) consists of \texttt{BloodVessel} elements (\fbox{BV}) for branches with parameters $\zparamb$ derived purely from the \gls*{3D} geometry. The \textit{optimized} 0D model (right) includes additional \texttt{BloodVesselJunction} elements (\fbox{BVJ}) for junctions and all parameters $\zparamo$ are optimized from \gls*{3D} simulation results.\label{fig_overview_models}}
\end{figure}

\glsresetall
Subject-specific \gls*{CFD} is an emerging tool in the treatment of cardiovascular diseases \cite{schwarz23}. Such \gls*{CFD} models require solving an initial boundary value problem where the domain of interest is commonly modeled with high-fidelity \gls*{3D} models while \gls*{0D} \glspl*{BC} physiologically approximate the downstream vasculature (Figure~\ref{fig_overview_models}, left). A common choice for \glspl*{BC} is the three-element Windkessel model \cite{vignon-clementelOutflowBoundaryConditions2006}, which approximates the vascular resistance and compliance of distal vascular beds that are not resolved in image data and computationally intractable in \gls*{3D} simulations. \gls*{0D} models (Figure~\ref{fig_overview_models}, middle and right) can estimate bulk blood flow and pressure at a fraction of the computational cost \cite{pfaller22} of \gls*{3D} simulations. In addition, \gls*{0D} models are used as efficient (albeit sometimes less accurate) substitutes for \gls*{3D} models in the simulation domain, e.g., to predict the pressure drop in aortic coarctations \cite{nair24} or coronary arteries \cite{schrauwen14}. They are especially useful in combination with \gls*{3D} models in many-query settings, such as uncertainty quantification or optimization problems involving thousands of model evaluations \cite{garber21}.

Windkessel \gls*{BC} parameters are patient-specific, meaning they must be calibrated for each patient based on clinical measurements \cite{vignon-clementelPrimerComputationalSimulation2010}, typically from 4D flow \gls*{mri} or  Doppler ultrasound flow measurements and catheter or cuff pressure measurements. In the following, these clinical measurements are referred to as observations. These observations are usually corrupted by noise, for example, from the limited accuracy of a measurement device or the unpredictable positioning of a sensor. \Glspl*{BC} estimated based on these measurements are thus more accurately represented by a probability distribution over a range of possible values rather than discrete parameter values. The benefit of quantifying these inherent uncertainties is twofold. First, it provides a measure of confidence when selecting a discrete point estimate of \glspl*{BC} for further analysis. Second, we can further propagate the uncertainties in \glspl*{BC} to uncertainties in model predictions for clinical decision-making. In this work, we infer the posterior distribution of \gls*{BC} parameters given noisy clinical observations. Our algorithm requires only minimal \gls*{3D} simulation costs, i.e., only a single fully resolved \gls*{3D} simulation over one cardiac cycle. To achieve this, we propose two novel techniques. First, we present a fast deterministic optimization method to yield a highly accurate \gls*{0D} model from \gls*{3D} data. Second, we use the \gls*{0D} model to predict the sought posterior distribution of the \gls*{BC} parameters given noisy observations.

The accuracy of \gls*{0D} models compared to \gls*{3D} fluid dynamics has been extensively quantified in previous studies \cite{schrauwen14,chnafa17,mirramezani20,pewowaruk21,pfaller22}. Automatic or semi-automatic tools generate these \gls*{0D} models from \gls*{3D} geometries, which typically achieve maximum approximation errors of bulk flow and pressure at their outlets of around 10\%, depending on the anatomy and the error metric used. In some examples, resistances in \gls*{0D} models have been calibrated semi-automatically or automatically to improve agreement between \gls*{0D} and \gls*{3D} simulations. For pulmonary models, Lan et al.~\cite{lan22} used a derivative-free Nelder-Mead optimization, whereas Lee et al.~\cite{lee23} used iterative linear corrections. Tran et al. \cite{tran17} iteratively refined scaling global factors for key coronary model outputs. In this work, we propose a deterministic least-squares procedure to optimize a \gls*{0D} model to match an initial \gls*{3D} model. Our algorithm optimizes all \gls*{0D} parameters in the model, i.e., resistances, inductances, capacitances, and non-linear stenosis factors. Our \gls*{0D} optimization method approximates the given data in a least-squares sense, which makes it applicable to arbitrary vascular anatomies.  In this work, we demonstrate its use in aorta, aortafemoral, coronary, and pulmonary models.

We refer to Marsden \cite{marsden14} for a general overview of optimization in cardiovascular modeling and to Ninos et al. \cite{ninos21} for an overview of \gls*{uq} in hemodynamics. Furthermore, we point out efficient developments in multi-fidelity uncertainty quantification \cite{nitzler2022generalized, biehler2015towards, biehler2019multifidelity, koutsourelakis2009accurate}, which are especially useful for large-scale coupled problems, and can drastically accelerate the otherwise expensive calculations. Different strategies have been demonstrated to calibrate the parameters of Windkessel \glspl*{BC} \cite{nolte22}. Spilker et al. \cite{spilkerTuningMultidomainHemodynamic2010} estimated Windkessel parameters by solving a system of non-linear equations using a quasi-Newton method informed by a \gls*{0D} surrogate to compute the initial Jacobian. Ismail et al. \cite{ismailAdjointbasedInverseAnalysis2013} proposed an adjoint-based approach to compute the initial Jacobian, thereby reducing the sensitivity to accurate initial guesses. Tran et al. \cite{tran17} and Menon et al. \cite{Menon2023,menon23} used derivative-free optimization to estimate high-dimensional Windkessel as well as other 0D BC parameters for coronary hemodynamics models informed by clinical measurements of cardiac function and coronary perfusion. However, deterministic optimization is prone to get stuck in unfavorable local optima, especially in the presence of experimental noise \cite{nolte22}. More robust approaches are based on \textit{Bayesian} calibration, which interprets the calibration as a statistical inference task, treating all variables as distributions. Here, the solution is captured by a probability distribution for the parameters. Pant et al. \cite{pantMethodologicalParadigmPatientspecific2014} demonstrated the application of Kalman filters for Windkessel calibration using an unscented Kalman filter (UKF). The UKF was applied to a \gls*{0D} surrogate of the \gls*{3D} model, which itself was subsequently recalibrated based on \gls*{3D} simulation results. Kalman filters sequentially incorporate observations about a dynamic process (like time-dependent clinical measurements) by propagating the mean and variance of the current parameter estimate and correcting the latter based on the observations. The UKF is a special form of Kalman filter, where dedicated control points are chosen for the forward propagation to reduce the number of required model evaluations \cite{pantMethodologicalParadigmPatientspecific2014}. Nevertheless, Kalman filters can only approximate Gaussian distributions. It has been shown, however, that posterior distributions in cardiovascular \glspl*{BC} deviate from Gaussian distributions, sometimes even exhibiting multiple modes \cite{spitierisBayesianCalibrationArterial2022}. Other methods use multi-fidelity approaches that rely on iteratively sampling the parameter space and learning a relationship between high-fidelity \gls*{3D} models and low-fidelity \gls*{0D} or \gls*{1D} models \cite{seo20,fleeter20}.

In this work, we propose a new framework for calibrating Windkessel parameters in cardiovascular models using \gls*{SMC}. As in Kalman filter approaches, the calibration problem is stated as a Bayesian inference task, and the result is given as the full posterior distribution over \gls*{BC} parameters given the observations. The probabilistic approach also incorporates prior knowledge about the \gls*{BC} parameters and information about the noise structure in, e.g., clinical measurements used for the calibration. \gls*{SMC} is employed to solve the Bayesian calibration problem. \gls*{SMC} can capture arbitrary posteriors, including multimodal distributions. By exploring intermediate distributions between the prior and the posterior, sampling is sequentially refined in regions with high posterior density, leading to computational savings. In our work, the workload of the Bayesian calibration is completely shifted to \gls*{0D} models. The \gls*{0D} models are deterministically optimized to a single \gls*{3D} simulation, enabling the Bayesian calibration's accuracy. This yields highly accurate \gls*{0D} models in unprecedented agreement with the corresponding high-fidelity \gls*{3D} models. 

The remainder of this work is structured as follows: In Section~\ref{sec_methods}, we outline the basic concepts for \gls*{3D} and \gls*{0D} modeling and the optimization and Bayesian calibration methods employed in this work. Section \ref{sec_results} presents the results of the \gls*{0D} model optimization and \gls*{BC} calibration algorithms and validates the results against the high-fidelity \gls*{3D} model. Section~\ref{sec_conclusion} summarizes the results of this work and provides an outlook for future work.

\section{Methods\label{sec_methods}}
\glsresetall
In this section, we introduce Bayesian Windkessel calibration (\ref{sec_multi_fidelity}).
Further, we briefly review governing equations for high-fidelity \gls*{3D} (\ref{sec_3D}) and surrogate \gls*{0D} (\ref{sec_0D}) blood flow models. We then extend these to the inverse problem of optimizing \gls*{0D} models  (\ref{sec_0D_optimization}).

\subsection{Bayesian Windkessel calibration with optimized 0D models\label{sec_multi_fidelity}}

The goal of the Bayesian \gls*{BC} calibration is to infer a posterior distribution $\probz(\param|\yobs,\zparam)$ for the Windkessel \gls*{BC} parameters $\param$, informed by uncertain clinical observations $\yobs$. We perform this using a \gls*{0D} model $\modelz$ with internal \gls*{0D} element parameters $\zparam$. The distinction between the Windkessel \gls*{BC} parameters $\param$ and the \gls*{0D} element parameters $\zparam$ is crucial for the following discussion and will be explained in detail below. In the following, we introduce probabilistic calibration (\ref{sec_bayesian_calibration}), describe the likelihood model selection in this work (\ref{sec_likelihood}), and present the \gls*{SMC} algorithm (\ref{sec_smc}).

\subsubsection{Bayesian calibration \label{sec_bayesian_calibration}}
The posterior distribution reflects the solution of the calibration and can be derived using Bayes' rule \cite{bertsekasIntroductionProbability2002}
\begin{equation}
    \posterior = \frac{\prob(\yobs|\param) \cdot \prob(\param)}{\prob(\yobs)}.
\end{equation}
The \textit{prior distribution} $\prob(\param)$ encodes the knowledge about the parameters before new data is observed. The \textit{likelihood} $\prob(\yobs|\param)$ defines a measure of proximity between the model response $\boldsymbol{y}=\modelz(\param)$ and the observations $\yobs$. The \textit{evidence} $\prob(\yobs)$ normalizes the former two components to an actual probability density. This normalization constant is typically difficult to evaluate, requiring the potentially high-dimensional integration over $\param$. \reva{To avoid this complexity, we take the same approach as many other algorithms by using the \textit{unnormalized posterior}}
\begin{equation}
\prob(\param|\yobs) \propto \prob(\yobs|\param) \cdot \prob(\param),
\end{equation}
\reva{and the logarithmic version of the latter and conduct the normalization \textit{a posteriori}.}

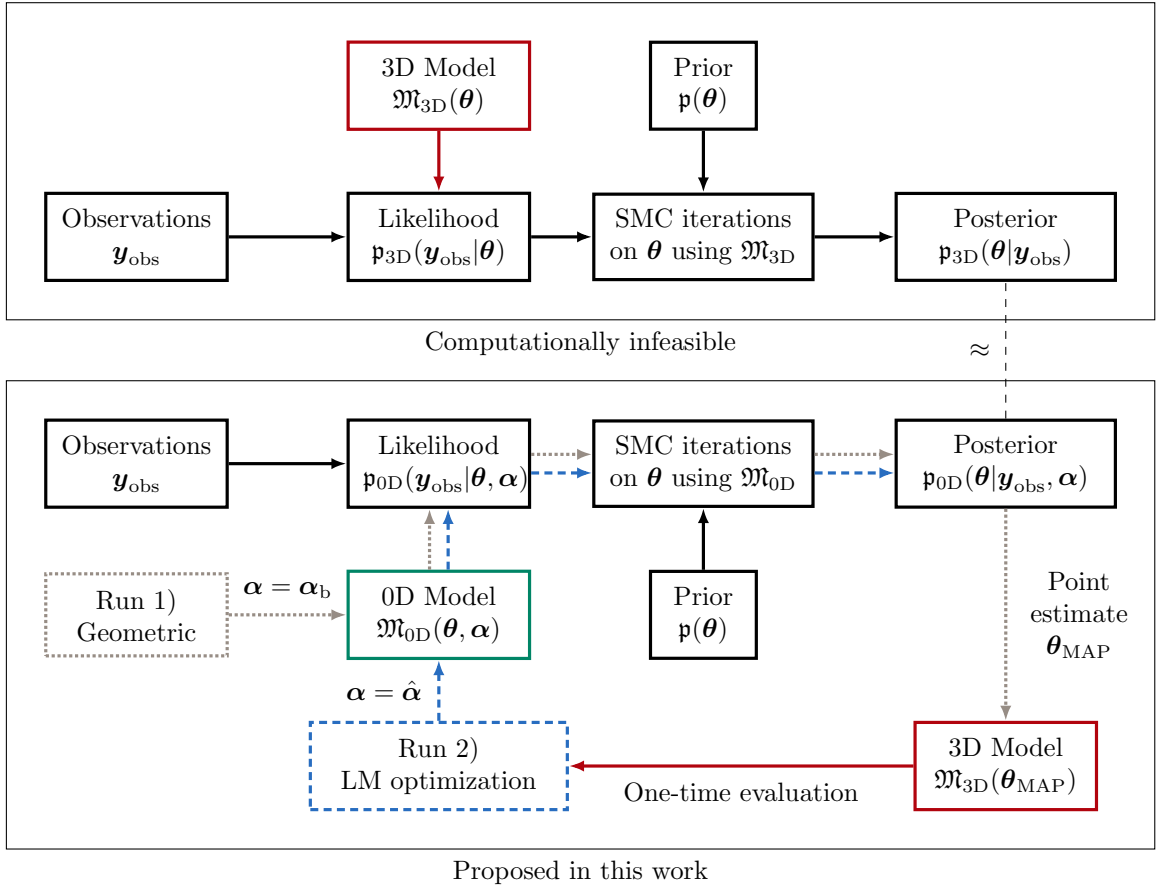
\begin{figure}[H]
\centering
\begin{tikzpicture}[auto, >={Latex[length=2mm]}]
\tikzset{every
node/.style={inner sep=0.2cm},
block/.style={rectangle, draw, text width=2cm, align=center, minimum height=1cm, very thick},
group/.style={rectangle, draw, inner sep=.5cm},
title/.style={align=center},
colorslow/.style={draw=clslow, very thick},
colorfast/.style={draw=clfast, very thick},
coloroptimized/.style={draw=cloptimized, very thick, densely dashed},
colorgeometric/.style={draw=clgeometric, very thick, densely dotted}
}
%
%
\node [block] (obs3) [text width=2cm, align=center]{Observations\\$\yobs$};
\node [block, right of=obs3, xshift=3cm] (likely3) [text width=2cm, align=center]{Likelihood\\$\probt(\yobs|\param)$};
\node [block, right of=likely3, xshift=2.5cm] (smc3) [text width=2.5cm, align=center]{SMC iterations\\on $\param$ using $\modelt$};
\node [block, colorslow, above of=likely3, yshift=1cm] (model3) [text width=2cm, align=center]{3D Model\\$\modelt(\param)$};
\node [block, above of=smc3, yshift=1cm] (prior3) [text width=1cm, align=center]{Prior\\$\prob(\param)$};
\node [block, right of=smc3, xshift=3cm] (post3) [text width=2.5cm, align=center]{Posterior\\$\probt(\param|\yobs)$};
\draw[->,very thick] (obs3) -- node {} (likely3);
\draw[->,very thick,colorslow] (model3) -- node {} (likely3);
\draw[->,very thick] (likely3) -- node {} (smc3);
\draw[->,very thick] (prior3) -- node {} (smc3);
\draw[->,very thick] (smc3) -- node {} (post3);
\node[group, fit=(obs3) (post3) (model3)] (box1) {};
\node[title] (title1) at ([yshift=-.3cm]box1.south) {Computationally infeasible};
%
%
\node [block, below of=obs3, yshift=-2cm] (obs0) [text width=2cm, align=center]{Observations\\$\yobs$};
\node [block, right of=obs0, xshift=3cm] (likely0) [text width=2cm, align=center]{Likelihood\\$\probz(\yobs|\param,\zparam)$};
\node [block, right of=likely0, xshift=2.5cm] (smc0) [text width=2.5cm, align=center]{SMC iterations\\on $\param$ using $\modelz$};
\node [block, colorfast, below of=likely0, yshift=-1cm] (model0) [text width=2cm, align=center]{0D Model\\$\modelz(\param,\zparam)$};
\node [block, colorgeometric, left of=model0, xshift=-3cm] (geo) {Run 1) Geometric};
\node [block, below of=smc0, yshift=-1cm] (prior0) [text width=1cm, align=center]{Prior\\$\prob(\param)$};
\node [block, right of=smc0, xshift=3cm] (post0) [text width=2.5cm, align=center]{Posterior\\$\probz(\param|\yobs,\zparam)$};
\node [block, colorslow, below of=post0, yshift=-3cm] (map3) {3D Model\\$\modelt(\paramm)$};
\node [block,coloroptimized, below of=model0, yshift=-1cm] (lm) [text width=3cm, align=center]{Run 2)\\LM optimization};
\draw[->,very thick,] (obs0) -- node {} (likely0);
\draw[->,very thick,colorgeometric,transform canvas={xshift=-3.5pt}] (model0) -- node {} (likely0);
\draw[->,very thick,coloroptimized,transform canvas={xshift=+3.5pt}] (model0) -- node {} (likely0);
\draw[->,very thick,coloroptimized,transform canvas={yshift=-3.5pt}] (likely0) -- node {} (smc0);
\draw[->,very thick,colorgeometric,transform canvas={yshift=+3.5pt}] (likely0) -- node {} (smc0);
\draw[->,very thick] (prior0) -- node {} (smc0);
\draw[->,very thick,coloroptimized,transform canvas={yshift=-3.5pt}] (smc0) -- node {} (post0);
\draw[->,very thick,colorgeometric,transform canvas={yshift=+3.5pt}] (smc0) -- node {} (post0);
\draw[->,very thick,colorgeometric] (post0) -- node [text width=1.5cm, align=center]{Point estimate\\$\param_\text{MAP}$} (map3);
\draw[->,very thick,colorslow] (map3) -- node {One-time evaluation} (lm);
\draw[->,very thick,coloroptimized] (lm) -- node {$\zparam=\zparamo$} (model0);
\draw[->,very thick,colorgeometric] (geo) -- node {$\zparam=\zparamb$} (model0);
\node[group, fit=(obs0) (post0) (geo) (lm)] (box2) {};
\node[title] (title2) at ([yshift=-.3cm]box2.south) {Proposed in this work};
\draw [dashed] (post0) -- node {$\approx$} (post3);
\end{tikzpicture}
\caption{Framework proposed in this work for Bayesian Windkessel \gls*{BC} calibration, using \gls*{3D} models $\modelt$ (expensive, red) and \gls*{0D} models (cheap, green). The \gls*{0D} models use geometric parameters (gray, \reva{dotted}) or optimized parameters (blue, \reva{dashed}).\label{fig_overview}}
\end{figure}

Bayesian calibration of Windkessel \gls*{BC} parameters $\param$ from observations $\yobs$ is a computationally challenging task. Using \gls*{SMC}, it easily requires thousands of model evaluations $\model$, each requiring a \gls*{3D} FEM solution (Figure~\ref{fig_overview_models}, left), to yield the posterior distribution $\posterior$ from a prior distribution $\prob(\param)$. Figure~\ref{fig_overview} (top) highlights this process using the \gls*{3D} model $\modelt$ (red) to obtain the posterior distribution $\probt(\param |\yobs)$. While this would give an exact representation of the posterior distribution, it is generally infeasible due to its excessive computational demand.

In this work, we combine models $\modelz$ and $\modelt$ in a novel calibration method visualized in Figure~\ref{fig_overview} (bottom). Here, expensive statistical model evaluation is performed exclusively with fast \gls*{0D} models $\modelz$ (green), relying on two crucial assumptions. First, we assume that we can construct highly accurate \gls*{0D} surrogate models that approximate $\modelt \approx \modelz$. Second, we assume that the \gls*{0D}-derived posterior approximates $\probt\approx\probz$. While \gls*{0D} models are orders of magnitude faster to evaluate, they are, in general, not sufficiently accurate to justify these assumptions. As part of our calibration process, we thus optimize the \gls*{0D} element parameters $\zparam$ from a single \gls*{3D} simulation to greatly enhance \gls*{0D} accuracy. This also allows us to demonstrate the validity of the above two assumptions.

We perform two \gls*{SMC} calibration runs in the proposed calibration workflow. The first \gls*{SMC} run (gray) is performed using a \textit{geometric} \gls*{0D} model (Figure~\ref{fig_overview_models}, middle) $\modelz(\param,\zparamb)$ composed of \texttt{BloodVessel} elements with parameters $\zparamb$ derived purely from the blood vessel's geometry, containing the vessel branches' resistances, inductances, capacitances, and non-linear stenosis factors (Section~\ref{sec_0D}) \cite{pfaller22}. This yields a first estimate for the posterior distribution $\probz(\param |\yobs,\zparamb)$. Using the parameter set $\param_\text{MAP}$ with the highest posterior density, the so-called \gls*{MAP} estimate, we perform one evaluation of the \gls*{3D} model $\modelt(\paramm)$ (Section~\ref{sec_3D}). These results are then used to determine the \textit{optimized} \gls*{0D} model (Figure~\ref{fig_overview_models}, right) parameters $\zparamo$, including \texttt{BloodVesselJunction} elements, using the \gls*{LM} algorithm (Section~\ref{sec_0D_optimization}). This optimization yields a highly accurate \gls*{0D} model to approximate $\modelt(\paramm) \approx \modelz(\paramm,\zparamo)$. Finally, we perform a second \gls*{SMC} (blue) run with the optimized \gls*{0D} model $\modelz(\param,\zparamo)$ to derive the final posterior distribution $\probz(\yobs|\param,\zparamo)$ for the Windkessel \gls*{BC} parameters $\param$. \revb{Note that the second \gls*{SMC} run is independent from the first one to not introduce any bias from the geometric \gls*{0D} model.} Throughout this manuscript, we refer to obtaining the optimal point estimate \gls*{0D} parameters $\zparamo$ using \gls*{LM} as \textit{optimization}. This is in contrast to the Bayesian \textit{calibration} of Windkessel parameters $\param$ with posterior distribution $\posterior$ under uncertain observations $\yobs$ using \gls*{SMC}.

\subsubsection{Zero-dimensional likelihood model\label{sec_likelihood}}

The forward model can only be evaluated pointwise, i.e., for a given set of parameters $\param$. The likelihood model scores the model output vector $\boldsymbol{y}=\modelz(\param)$ for a given parameter vector $\param$ based on the output's proximity to the observations $\yobs$. Moreover, it incorporates knowledge about the noise in the observations resulting from the clinical measurements \cite{kaipioStatisticalComputationalInverse2005}. In this work, an \textit{additive noise} model for the observations is chosen according to Kaipio et al. \cite{kaipioStatisticalComputationalInverse2005} as
\begin{equation}
    \yobs = \boldsymbol{y} + \boldsymbol{\varepsilon} =  \modelz(\param) + \boldsymbol{\varepsilon}.
\end{equation}
\reva{For simplicity,} the noise is assumed to be normally distributed with zero mean and covariance matrix $\boldsymbol{\Sigma}$, i.e., $\boldsymbol{\varepsilon}\sim \mathcal{N}(\boldsymbol{0}, \boldsymbol{\Sigma})$. The covariance matrix is further assumed to be diagonal wherein the variance vector $\boldsymbol{\sigma}^2$ denotes the diagonal entries of $\boldsymbol{\Sigma}$.  We choose the commonly used Gaussian likelihood model here, which is defined as:
\begin{equation}
    \prob\left(\yobs|\param\right)=\frac{1}{\sqrt{\left(2 \pi \right)^n  \mathrm{det}\, \boldsymbol{\Sigma}}} \exp \left[-\frac{1}{2} \left(\modelz(\param) - \yobs\right)^\top\boldsymbol{\Sigma}^{-1} \left( \modelz(\param) - \yobs\right)\right]
    \label{eq:likelihood}
\end{equation}
It should be noted that $\prob\left(\yobs|\modelz(\param)\right)$ is not a valid probability density in $\param$. Therefore, it is often referred to as \textit{likelihood function} in $\param$. \reva{The Gaussian likelihood maximizes information entropy and thus introduces the least bias in the absence of more detailed noise information. Nevertheless, our proposed framework works with any likelihood model and can incorporate specific measurement biases from clinical measurements, introducing additional noise parameters. For example, magnetic resonance magnitude images have been shown to exhibit a Rice noise distribution \cite{gudbjartsson95}.}

\subsubsection{Sequential Monte Carlo\label{sec_smc}}
Different methods have been developed to solve the Bayesian calibration problem. We employ the \gls*{SMC} algorithm, also known as particle filters \cite{chopinIntroductionSequentialMonte2020, chopinSequentialParticleFilter2002}, which is described in more detail in Appendix~\ref{sec_smc_details}. The posterior distribution $\pi$ in \gls*{SMC} is approximated by weighted particles according to
\begin{equation}
    \prob(\param|\yobs) \approx \sum_{i=1}^k W_{n+1}^{(i)} \, \delta_{\param_{n+1}^{(i)}}(\param),
\end{equation}
with normalized weights $W$, number of particles $k$, and final iteration number $n$. Each particle $\delta_{\param_n^{(i)}}$ refers to a Dirac delta distribution at particle location $\param_i$. Using the parameter vector $\param$ with the highest posterior probability \cite{kaipioStatisticalComputationalInverse2005}, the so-called \gls*{MAP} estimate can be computed as
\begin{equation}
    \param_{\text{MAP}} = \mathrm{arg} \max_{\param} \left[\, \prob(\param|\yobs) \,\right].
\end{equation}
In \gls*{SMC}, this is equivalent to selecting the particle location $\param_i$ of the particle with the highest weight.

\subsection{Three-dimensional blood flow modeling\label{sec_3D}}
The \gls*{3D} model $\modelt$ models blood as a Newtonian fluid in the rigid fluid domain $\Omega^\text{3D}$ over time $T$ (typically several cardiac cycles) governed by the incompressible Navier-Stokes equations, 
\begin{alignat}{3}
\nabla \cdot \bs \sigma(\vc u, p) &= \rho \left[ \dot{\vc u} + (\vc u \cdot \nabla) \vc u \right] , \quad && \vc{x} \in \Omega^\text{3D}, \quad t \in [0,T], \label{eq_3D_momentum}\\
\nabla \cdot \vc u &= 0, \quad && \vc{x} \in \Omega^\text{3D}, \quad t \in [0,T], \label{eq_3D_continuity}
\end{alignat}
with velocity $\vc u$, pressure $p$, Cauchy stress tensor $\bs \sigma = \mu(\nabla\vc u + \nabla\vc u^{\mathrm{T}})-p\vc I$, constant density $\rho$, and constant dynamic viscosity $\mu$. Throughout this manuscript, $\dot{\phantom{m}}$ denotes a time derivative. We prescribe \glspl*{BC}
\begin{alignat}{4}
\vc{u}(\vc{x}, t) &= \vc{0}, \quad && \vc{x} \in \Gamma_\text{wall}, \quad && t \in [0,T],\\
\vc{u}(\vc{x}, t) &= \vc{u}_\text{in}(\vc{x},t), \quad && \vc{x} \in \Gamma_\text{in}, \quad && t \in [0,T], \label{eq_3D_BC_in}\\
\vc{f}_\text{WK3} \left(P_i(t), Q_i(t), \,\param_i \right) &= \vc 0, \quad && \vc{x} \in \Gamma_{\text{out},i}, \quad && t \in [0,T], \label{eq_3D_BC_out}\\
\text{with~} p(\vc x, t) &= P_i(t),\\
 Q_i(t) &= \int_{\Gamma_{\text{out},i}} \vc{u}(\vc{x}, t) &&\cdot \vc n \,\text{d}\vc x,
\end{alignat}
with a no-slip \gls*{BC} at the vessel wall $\Gamma_\text{wall}$, time-dependent parabolic inflow profile $\vc{u}_\text{in}$ at the inlet $\Gamma_\text{in}$. At each outlet $\Gamma_{\text{out},i}$, we prescribe a three-element Windessel \gls*{BC} with parameters $\param_i$ that relate flow rate $Q_i$ and the pressure $P_i$ at outlet $i$ through an ordinary differential equation $\vc{f}_\text{WK3}$. Throughout this work, we use the three-element Windkessel model defined in Appendix~\ref{sec_0D_ele_windkessel} for all outlets. We note that this choice does not restrict the \gls*{BC} calibration approach proposed in this work as it applies generally to any \gls*{BC} type. For details on the \gls*{0D}-\gls*{3D} coupling approach, we refer to Vignon-Clementel et~al. \cite{vignonclementel06, vignonclementel10,moghadam13}. The initial conditions are
\begin{alignat}{3}
\vc{u}(\vc{x}, t=0) &= \vc{u}_0(\vc{x}), \qquad && \vc{x} \in \Omega^\text{3D}\\
p(\vc{x}, t=0) &= p_0(\vc{x}), \qquad && \vc{x} \in \Omega^\text{3D}
\end{alignat}
with initial velocity field $\vc{u}_0$ and initial pressure field $p_0$, which we estimate by projecting a periodic \gls*{0D} solution to the \gls*{3D} geometry \cite{pfaller21}.

\subsection{Zero-dimensional blood flow modeling}
\label{sec_0D}
Analogous to electrical circuits, the elemental  building blocks for blood vessels of the \gls*{0D} model $\modelz$ are defined as
\begin{align}
\Delta P = \od{R} Q, \quad \Delta P = \od{L} \dot{Q}, \quad Q = \od{C} \dot{P}, \quad \Delta P = \od{S} Q|Q|,
\label{eq_0D_blocks}
\end{align}
with \gls*{0D} parameters resistance $\od{R}$, inductance $\od{L}$, capacitance $\od{C}$, and stenosis coefficient $\od{S}$. Note that only the stenosis term depends non-linearly on the flow $Q$. The operator, $\Delta\bullet$ characterizes the difference between element \reva{inlet $\bullet_\text{in}$} and \reva{outlet $\bullet_\text{out}$}.

We use the \reva{methods} outlined in Pfaller~et~al.~\cite{pfaller22} to split the vessel into branches and junctions \reva{and determine values for $\mathcal{RLCS}$ elements \eqref{eq_0D_blocks}, which we} combine into a \texttt{BloodVessel} \gls*{LPN} element as described in Appendix~\ref{sec_0D_ele_vessel}. \reva{In short,} we predict the branch parameter values $\zparamb$ of each \texttt{BloodVessel} element as
\begin{align}
\reva{\mathcal{R} = \frac{8\mu l}{\pi r^4}, \quad \mathcal{L} = \frac{\rho l}{\pi r^2}, \quad \mathcal{C} = \frac{3l\pi r^3}{2Eh}, \quad \mathcal{S} = K_t \frac{\rho}{2S_0^2} \left( \frac{S_0}{S_s} - 1 \right)^2,}
\label{eq_0D_values}
\end{align}
\reva{with dynamic viscosity $\mu$ and density $\rho$ of blood, lumen radius $r$, blood vessel length $l$, and vessel wall Young's modulus $E$ and thickness $h$. The stenosis factor $\mathcal{S}$ includes the stenosed area $S_s$, proximal cross-sectional area $S_0$, and empirical correction factor $K_t=1.52$ \cite{steele03}. The areas $S_s$ and $S_0$ and the choice to split the branch into three segments are based on the relative extrema of the branch's cross-sectional area.}

In the following sections, these models are referred to as \textit{geometric} \gls*{0D} models. We also define a \texttt{BloodVesselJunction} \gls*{LPN} element to connect an inlet to two or more outlets using the $\mathcal{RLS}$ elements with parameters $\zparamj$, as described in Appendix~\ref{sec_0D_ele_junction}. The \texttt{BloodVesselJunction} \gls*{LPN} element does not include a capacitance, which facilitates the derivation of the Jacobian for the zero-dimensional model optimization. Note that we do not currently infer \texttt{BloodVesselJunction} element parameters from the blood vessel geometry. Predicting junction parameters from the vessel geometry remains an active research topic \cite{mynard15,rubio24}. Thus, all junction parameters $\zparamj$ are set to zero in \textit{geometric} \gls*{0D} models, i.e., $\zparamj=\vc0$. Nevertheless, we optimize parameters for both \texttt{BloodVessel} and \texttt{BloodVesselJunction} elements in Section~\ref{sec_0D_optimization} using \gls*{3D} simulation results. These models are referred to as \textit{optimized} \gls*{0D} models.

Connecting all elements $e$ within a vascular tree forms a \gls*{LPN} with parameters $\zparam=\zparamb\cup\zparamj$. As in the \gls*{3D} model (Section~\ref{sec_3D}), we prescribe the flow rate $Q$ at the inlet and couple the outlets to Windkessel \glspl*{BC}. We assemble all elements $e$ and \glspl*{BC} in our \gls*{LPN} into a global residual vector $\vc r$. The global system of differential-algebraic equations can be split by the dependence on the solution vector $\vc y$, \revb{containing pressures and flow-rates at \gls*{LPN} node points}, and its time derivative $\dot{\vc y}$ as
\begin{align}
\vc r (\zparam, \vc y, \dot{\vc y}) = \vc E (\zparam) \cdot \dot{\vc y} + \mathbf{F} (\zparam)  \cdot \vc y + \vc c (\zparam, \vc y, \dot{\vc y}),
\label{eq_0D_residual}
\end{align}
with constant parameter-dependent system matrices $\vc E$ and $\vc F$ and system vector $\vc c$ that depends non-linearly on $\vc y$ and $\dot{\vc y}$. We compute pressure and flow rate associated with a given set of \gls*{0D} parameters $\zparam$ by solving
\begin{align}
\text{Solve~}\vc r (\zparam, \vc y, \dot{\vc y}) \overset{!}{=} \vc 0 \text{~for~} \vc y, \dot{\vc y} \text{~with given~} \zparam, \quad \revb{\text{in}~\Omega^\text{0D}}, \quad t \in [0,T], \label{eq_0D_forward}
\end{align}
which, in the following, is referred to as the \textit{forward} \gls*{0D} problem. The node points of the LPN, i.e., where \gls*{0D} elements connect, make up the domain $\Omega^\text{0D}$. We generate maps $\Omega^\text{0D}\to\Omega^\text{3D}$ and $\Omega^\text{3D}\to\Omega^\text{0D}$ through a region-growing algorithm and centerline integration, respectively, to map results between \gls*{0D} and \gls*{3D} models \cite{pfaller21}. We prescribe \glspl*{BC} at the inlet and outlets
\begin{alignat}{4}
Q_j(t) &= Q_\text{in}(t), \quad && \revb{\text{on} ~ \Gamma_\text{in}}, \quad && t \in [0,T],\\
\vc{f}_\text{WK3} \left(P_j(t), Q_j(t), \,\param_j \right) &= \vc 0, \quad && \revb{\text{on} ~ \Gamma_{\text{out},i}}, \quad && t \in [0,T],
\end{alignat}
with inflow rate $Q_\text{in}$ and three-element Windkessel \glspl*{BC} $\vc{f}_\text{WK3}$ corresponding to the \gls*{3D} model inlet and outlet \glspl*{BC} $\param$, \eqref{eq_3D_BC_in} and \eqref{eq_3D_BC_out}, respectively. Furthermore, we prescribe initial conditions
\begin{align}
\vc y(t=0) = \bar{\vc y},
\end{align}
where we obtain $\bar{\vc y}$ from a steady-state simulation with mean inflow and steady-state \glspl*{BC}. This lowers the number of cardiac cycles required to reach a periodic solution compared to initializing with $\vc y(t=0) = \vc 0$. We use the implicit generalized-$\alpha$ method to discretize the forward problem~\eqref{eq_0D_forward} in time and the Newton-Raphson method to solve for the solution vector $\vc y$ in each time step. The details of the method are given in Appendix~\ref{sec_gen_alpha}. Reusing terms from the residual \eqref{eq_0D_residual}, the Jacobian matrix $\vc K$ for the time-discrete forward problem \eqref{eq_0D_forward} is computed as
\begin{align}
\vc K
= \left. \pfrac{\vc r}{\dot{\vc y}} \right|_{\zparam = \text{const}} 
= \alpha_m \left( \vc E + \pfrac{\vc c}{\dot{\vc y}} \right) + \alpha_f\gamma\Delta t_n \left( \vc F + \pfrac{\vc c}{\vc y} \right),
\end{align}
with time step $\Delta t_n$ and time integration parameters $\alpha_f$, $\alpha_m$, and $\gamma$. All system matrices for the \texttt{BloodVessel}, \texttt{BloodVesselJunction}, and \texttt{Windkessel} elements can be derived analytically and are provided in Appendices~\ref{sec_0D_ele_vessel}, \ref{sec_0D_ele_junction}, and \ref{sec_0D_ele_windkessel}, respectively.

\subsection{Zero-dimensional model optimization\label{sec_0D_optimization}}
The proposed method for Bayesian Windkessel calibration relies on highly accurate \gls*{0D} models $\modelz$ that act as a surrogate for a given \gls*{3D} model $\modelt$. In general, \gls*{0D} models with geometric parameters $\zparamb$ are insufficiently accurate. In this section, we describe how to derive an optimized \gls*{0D} parameter set $\zparamo$ that better approximates $\modelt(\paramm)\approx\modelz(\paramm,\zparamo)$ for a given set of Windkessel parameters $\param_\text{MAP}$. Solving an inverse problem, we infer \gls*{0D} model parameters $\zparamo$ that optimally (in the least-squares sense) approximate given \gls*{0D} solution vectors $\vc y$ and $\dot{\vc y}$. In this work, $\vc y$ is extracted from a \gls*{3D} model $\modelt$ evaluated at the parameter set $\param_\text{MAP}$. However, other sources like \textit{in vivo} or \textit{in vitro} measurements are also possible. The time derivative $\dot{\vc y}$ can be either taken from $\modelt$ or approximated from $\vc y$. We use the same residual~$\vc r$ from~\eqref{eq_0D_residual} as in the forward problem~\eqref{eq_0D_forward}:
\begin{align}
\text{Solve~} \vc r (\zparam, \vc y, \dot{\vc y}) \overset{!}{=} \vc 0 \text{~for~} \zparam \text{~with given~} \vc y, \dot{\vc y}, \quad \revb{\text{in}~\Omega^\text{0D}}.
\label{eq_0D_inverse}
\end{align}
To solve the inverse problem, we formulate \eqref{eq_0D_inverse} as a least-squares problem
\begin{align}
\zparamo = \arg \min_{\zparam} S, \quad\text{with~} S = \sum_j^{D} r_j^2(\zparam, y_j, \dot{y}_j),
\label{eq_0D_least_squares}
\end{align}
where $\zparamo$ is the optimal parameter set that minimizes the sum of squared residuals $S$, and $D$ is the number of observations. Here, we solve for both branch and junction \gls*{0D} parameters $\zparamob$ and $\zparamoj$, respectively. The value $D$ is typically given from the number of \gls*{LPN} node points times the number of observed time steps. \reva{Note that in case of sparse observations, the parameters $\hat\zparam$ might not be deterministically identifiable.} We use the \gls*{LM} algorithm to solve the least squares problem \eqref{eq_0D_least_squares} iteratively for the optimal parameter set $\zparamo$, which is described in more detail in Appendix~\ref{sec_optimization}. We calculate the Jacobian for each \gls*{0D} element as the gradient of the residual $\vc r$ with respect to the \gls*{0D} parameters $\zparam$,
\begin{align}
\vc J = \left. \pfrac{\vc r}{\zparam} \right|_{\vc y, \dot{\vc y} = \text{const}}
= \pfrac{\mathbf{E}}{\zparam} \cdot \dot{\vc y}
+ \pfrac{\vc F}{\zparam} \cdot \vc y 
+ \pfrac{\vc c}{\zparam}.
\label{eq_0D_jac}
\end{align}
Note that the solution vectors $\vc y, \dot{\vc y}$ are constant in \eqref{eq_0D_jac}, and the Jacobian can, again, be computed analytically. These terms are straightforward to derive and are given for the \texttt{BloodVessel} and \texttt{BloodVesselJunction} \gls*{LPN} elements in Appendices~\ref{sec_0D_ele_vessel} and \ref{sec_0D_ele_junction}, respectively. The resulting \gls*{0D} models $\modelz$ with optimal parameters $\zparamo=\zparamob\cup\zparamoj$ are denoted as \textit{optimized} \gls*{0D} models.

\subsection{Open-source software}
All results in this work were computed with open-source software. The \gls*{3D} simulations were computed in SimVascular's finite element solver \texttt{svSolver} \cite{svsolver}; see Esmaily Moghadam et~al. \cite{moghadam13,esmailymoghadam13} for details. We further developed \texttt{svZeroDSolver} \cite{svzerodsolver}, containing the algorithmic implementation of the methods described in \ref{sec_0D} and \ref{sec_0D_optimization}. The forward~\eqref{eq_0D_forward} and inverse problems~\eqref{eq_0D_inverse} use the same residual vector $\vc r$ \eqref{eq_0D_residual} with linearizations with respect to the solution vector $\vc y$ and the parameter vector $\zparam$, respectively. The forward problem requires time discretization, whereas the inverse problem is simultaneously solved for all time steps. Nevertheless, the shared residual allows for reuse of large parts of the code for both problems. The system matrices for all \gls*{0D} elements for forward and inverse problems are given in Appendix~\ref{sec_matrices}. The core functionality is implemented in C\texttt{++} and based on the fast linear algebra package \texttt{Eigen}. Python and C\texttt{++} interfaces enable seamless integration into other software \cite{lee23,menon23}. The calibration framework \texttt{svSuperEstimator} \cite{SvSuperEstimator2022} implements the Windkessel calibration described in Section~\ref{sec_multi_fidelity} and interfaces different tools for performing \gls*{0D} and \gls*{3D} simulations, postprocessing routines and \gls*{SMC} Windkessel calibration. The \gls*{SMC} implementation is based on the Python package \texttt{particles} \cite{particles}. All 72 vascular models are publicly available from the Vascular Model Repository \cite{vmr}.

\section{Computational results and discussion\label{sec_results}}
\glsresetall
Our results showcase the two-run process of Bayesian calibration of Windkessel \gls*{BC} parameters $\param$ given uncertain observations $\yobs$. First, Section~\ref{sec_res_0D_opti} demonstrates the approximation of \gls*{3D} models by \gls*{0D} models $\modelt(\param)\approx\modelz(\param,\zparamo)$ with optimized \gls*{0D} parameters $\zparamo$. Second, Section~\ref{sec_bc_calibration} provides posterior distributions for \gls*{BC} parameters $\param$ derived with \gls*{SMC} using the optimized \gls*{0D} models,  $\probt(\param|\yobs)\approx\probz(\param|\yobs,\zparamo)$.

\subsection{Optimization of zero-dimensional models\label{sec_res_0D_opti}}
In this section, we investigate the \gls*{LM} \gls*{0D} model optimization performance with the algorithm presented in Section \ref{sec_0D_optimization}. The \gls*{LM} optimizer derives new blood vessel and junction parameters $\zparamo$ from a \gls*{3D} simulation result, enabling the approximation $\modelt(\param)\approx\modelz(\param,\zparamo)$. The \gls*{0D} models are improved in two aspects: enhancing the geometrically derived \texttt{BloodVessel} elements $\zparamob$, and identifying suitable parameters for \texttt{BloodVesselJunction} elements $\zparamoj$ (which are zero in geometric \gls*{0D} models). We obtain the time derivative $\dot{\vc y}$ by fitting a periodic cubic spline to $\vc y$, taking its derivative, and resampling the solution to 100 time points. The number of data points $D$ was thus 100 times the number of \gls*{LPN} node points, which ranged from 28 in aortas to 1460 in pulmonary arteries (mean 274).

To demonstrate the robustness of the \gls*{0D} model optimization, we demonstrate the results on a large dataset of 72 vascular models, each with given Windkessel \gls*{BC} parameters $\paramvmr$ that were calibrated in prior studies. This openly available dataset \cite{vmr} is taken from Pfaller et al. \cite{pfaller22} and visualized in Figure~\ref{fig_collage}. It includes diverse vascular anatomies, including aortas, aortofemoral, pulmonary, and coronary arteries. Vascular states include normal, (artificial) aortic coarctation, Kawasaki disease, single ventricle disease, Marfan syndrome, abdominal aortic aneurysm, pulmonary artery hypertension, and others. Note that the \gls*{3D} dataset has rigid walls, corresponding to a theoretical \gls*{0D} capacitance of $\od{C}=0$. However, we found that using a small non-zero value for $\od{C}$, thus modeling a nearly rigid wall, is beneficial for \gls*{0D} numerical performance \cite{pfaller22}. For completeness, we still report results for $\od{C}$ in this section despite it not having a physical meaning in our \gls*{0D} models.

We first verify in Section~\ref{sec_lm_truth} that we can identify \gls*{0D} branch parameters $\zparamob$ from an optimization given ground truth \gls*{0D} models with geometric branch parameters $\zparamb$. We then demonstrate in Section~\ref{sec_lm_training} the  accuracy of \gls*{0D} models $\modelz(\paramvmr,\zparamo)$ with parameters $\zparamo$ optimized from \gls*{3D} models $\modelt(\paramvmr)$. Finally, we verify in Section~\ref{sec_lm_validation} that the optimized \gls*{0D} models generalize well to a wide range of \gls*{BC} parameters $\param$. These results confirm that we can confidently use optimized \gls*{0D} models for the approximation $\modelt(\param)\approx\modelz(\param,\zparamo)$ in the Bayesian Windkessel calibration in Section~\ref{sec_bc_calibration}.

\subsubsection{Identifying zero-dimensional parameters from ground truth data\label{sec_lm_truth}}
As a first step in assessing the \gls*{0D} element optimization proposed in Section~\ref{sec_0D_optimization}, we verify that our optimization can successfully recover ground truth \gls*{0D} geometric branch parameters $\zparamb$ from $\modelz$ data:
\begin{align}
\text{Compare~} \zparamob \text{~vs.~} \zparamb \text{~with~} \zparamob = \arg \min_{\zparam} \Vert \vc r (\zparam, \modelz(\paramvmr,\zparamb)) \Vert^2.
\label{eq_0D_ground_truth}
\end{align}
In the absence of any 3D-0D model discrepancy, the optimization should be able to perfectly recover the ground truth \gls*{0D} parameters $\zparamob\approx\zparamb$. Recall from Section~\ref{sec_intro} that $\zparamj=\vc 0$ in all geometric models which results in optimized junction parameters $\zparamoj = 
\vc 0$. We ran forward models $\modelz$ for all 72 geometric models with Windkessel \glspl*{BC} $\paramvmr$ from the Vascular Model Repository to generate the ground truth data. We extracted the solution of pressures and flow rates and their time derivatives from each model at all node points of the \gls*{0D} \glspl*{LPN}. We initialized all \gls*{0D} parameters $\zparam = \vc 0$ to not introduce any bias to the optimized parameters $\zparamo$. Providing $\modelz(\paramvmr,\zparamb)$ as input, we performed \gls*{LM} optimization and obtained results for the optimized branch parameters $\zparamob$.

Due to the efficient C\texttt{++} implementation and fast Python interface of \texttt{svZeroDSolver}, running all 72 forward \gls*{0D} simulations and optimizations took only 47\,s and 101\,s on a single core, respectively. The optimization typically converged in less than ten iterations, although three coronary models stagnated at residual norms slightly above the tolerances defined in Appendix~\ref{sec_optimization}. This could be due to pressures and flow rates spanning different orders of magnitude in the aorta and coronary arteries, which could be prevented by normalization. We provide a detailed overview of all optimized parameters $\zparamob$ vs. ground truth geometric parameters $\zparamb$ in Figure~\ref{fig_optimized_0D} in Appendix~\ref{sec_detailed_opt_results}, showing $\od{R}, \od{L}, \od{C},$ and $\od{S}$ for every branch. The mean of the coefficients of determination averaged over all 72 models and their branches was  $R^2_\od{R}=R^2_\od{L},=R^2_\od{C}=1.0$. For the stenosis coefficient, this value was slightly reduced to $R^2_\od{S}=0.95$.

These results confirm that the four parameters of the \texttt{BloodVessel} element, $\od{R}, \od{L}, \od{C}$, and $\od{S}$ are identifiable from the solution vector $\vc y$ and its time derivative $\dot{\vc y}$. The slightly lower value $R^2_\od{S}$ of the stenosis coefficient indicates that it is more difficult to identify. However, it should be noted that many of the stenosis parameters $\od{S}$ in the geometric \gls*{0D} models are negligibly small and have little influence on the result, which makes them difficult to identify. The stenosis parameter is well identifiable in models with pathological stenoses, like in aortic coarctations.

\subsubsection{Optimizing zero-dimensional models to three-dimensional data\label{sec_lm_training}}
We demonstrated in \ref{sec_lm_truth} the excellent identifiability of \gls*{0D} parameters from ground truth data using the \gls*{LM} optimization. In this section, we quantify how well optimized \gls*{0D} models $\modelz$ approximate high-fidelity \gls*{3D} models $\modelt$:
\begin{align}
\text{Compare~} \modelt(\paramvmr) \text{~vs.~} \modelz(\paramvmr,\zparamo) \text{~with~} \zparamo = \arg \min_{\zparam} \Vert \vc r (\zparam, \modelt(\paramvmr)) \Vert^2.
\label{eq_0D_3D_training}
\end{align}
We extracted $\modelt$ velocity and pressure from the last cardiac cycle at all time steps with Windkessel \gls*{BC} parameters $\paramvmr$ given from the Vascular Model Repository. Using a previously proposed technique to extract \gls*{3D} results along the centerline of the blood vessels, we mapped the cross-section averaged flow rate and pressure from \gls*{3D} to \gls*{0D} \gls*{LPN} node points \cite{pfaller22,SvSuperEstimator2022}. We initialized the \gls*{0D} parameters with the values from the geometric model as $\zparam^0_\text{b} = \zparamb$ and $\zparam^0_\text{j} = \vc 0$ in branches and junctions, respectively. For model 0081\_0001, we excluded the stenosis coefficient from \gls*{LM} optimization as the 0D model with optimized stenosis coefficient did not converge during forward evaluations. The optimization of all 72 models took 79\,s in total on a single core.

Figure~\ref{fig_optimized_3D} in Appendix~\ref{sec_detailed_opt_results} provides a detailed correlation of all optimized \gls*{0D} parameters $\zparamob$ vs. the geometric ones $\zparamb$. The mean of the coefficients of determination averaged over all 72 models were $R^2_\od{R}= 0.68, R^2_\od{L} = 0.89, R^2_\od{C} = 0.23$, and $R^2_\od{S}=0.24$. This optimization also incorporates \gls*{0D} junction parameters $\zparamoj$. However, since $\zparamj = \vc 0$ in the geometric \gls*{0D} models, we cannot provide a correlation for these values. Since we aim to minimize the \gls*{0D} approximation errors by optimizing element parameters \eqref{eq_0D_3D_training}, we do not expect the coefficients of determination to be close to one as in Section~\ref{sec_lm_truth}. Rather, they indicate where the model generation process from a \gls*{3D} geometry can still be improved. Comparing the optimized parameters $\zparamob$ to the geometric ones $\zparamb$ in Figure~\ref{fig_optimized_3D}, there are some systematic differences. As expected, due to the rigid wall nature of the \gls*{3D} results, the capacitances $\od{C}$ are optimized as near zero. Furthermore, the optimized stenosis parameters $\od{S}$ are consistently higher, indicating that they are systematically underestimated in the geometric models. However, some pulmonary models with higher post-optimization errors than others exhibit non-zero capacitances $\od{C}$. Given that these models have many more junctions than others, this could indicate limitations in our \texttt{BloodVesselJunction} model in representing \gls*{3D} fluid dynamics in a junction.

\begin{figure}[H]
\centering
    \includegraphics[width=\linewidth]{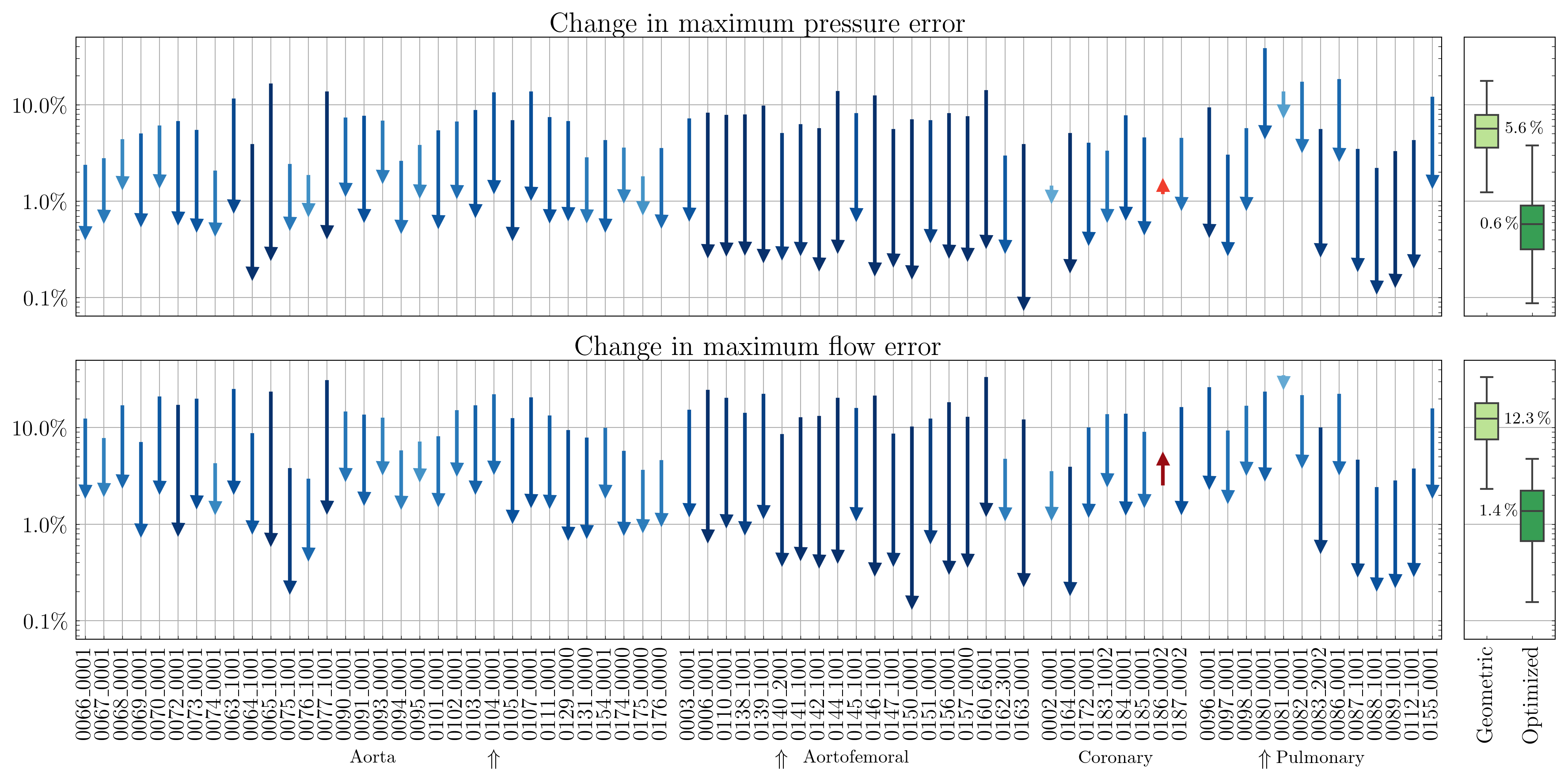}
\caption{Changes in maximum pressure and flow error between geometric and optimized \gls*{0D} models. Blue is an improvement; red is a deterioration. Darker values indicate larger changes. Statistical results are provided on the right. See Figure~\ref{fig_collage} for an overview of the geometries.
\label{fig_optimization_error}}
\end{figure}

Figure~\ref{fig_optimization_error} depicts the change in the errors between geometric \gls*{0D} models $\modelz(\paramvmr,\zparamb)$ and optimized \gls*{0D} models $\modelz(\paramvmr,\zparamo)$ with respect to \gls*{3D} models $\modelt(\paramvmr)$. The optimization improved the maximum approximation errors for pressures $\epsilon_{P,\text{max}}$ and flow rates $\epsilon_{Q,\text{max}}$ in 71 out of the 72 models, with error metrics defined as \cite{pfaller22}
\begin{equation}
\epsilon_{P,\text{max}} = \frac{n_t}{n_\text{cap}} \sum_{i=1}^{n_\text{cap}} \frac{\max_t \left| P_{t,i}^\text{0D} - P_{t,i}^{3\text{D}}\right|}{\sum_{t=1}^{n_t} P_{t,i}^\text{3D}}, \quad 
\epsilon_{Q,\text{max}} = \frac{1}{n_\text{out}} \sum_{i=1}^{n_\text{out}} \frac{\max_t \left| Q_{t,i}^\text{0D} - Q_{t,i}^{3\text{D}}\right|}{\max_t Q_{t,i}^\text{3D} - \min_t Q_{t,i}^\text{3D}},\\
\end{equation}   
with $n_t$ time steps $t$, number of outlets $n_\text{out}$, and number of caps $n_\text{cap} = n_\text{out} + 1$. Only the coronary model 0186\_0002 performed slightly worse after optimization. For this model, the junction between the aorta and the coronary arteries could not be calibrated properly. Overall, the median error was reduced from 5.6\% to 0.6\% for pressure and 12.3\% to 1.4\% for flow following optimization. The optimization achieved the largest improvements for aortofemoral models. Given that our approach optimizes all \gls*{0D} parameters, it achieves much lower approximation errors than previous studies where only resistances were adapted, e.g., to match key \gls*{3D} outputs within $10\%$ \cite{tran17}.

Optimizing the \gls*{0D} models to \gls*{3D} data lowered their maximum approximation error for pressure and flow rate by nearly one order of magnitude, from $\sim10\%$ to $\sim1\%$. This considerable improvement in \gls*{0D} accuracy is key to our Bayesian Windkessel calibration method. The improvements can be mainly attributed to the optimization of previously unknown \texttt{BloodVesselJunction} parameters $\zparamoj$. Since the geometric model does not capture the fluid behavior in the junctions, it consistently underestimates their pressure drop. Informing the \gls*{0D} model using a \gls*{3D} simulation result, it is easily possible to also derive suitable parameters for the junctions and thereby improve the \gls*{0D} model to model the full domain accurately. 

\subsubsection{Generalization of optimized zero-dimensional models\label{sec_lm_validation}}
We demonstrated in Section~\ref{sec_lm_training} that the \gls*{LM} optimization yields \gls*{0D} models $\modelz(\paramvmr,\zparamo)$ which reduce maximum pressure and flow errors with respect to \gls*{3D} models $\modelt(\paramvmr)$ by nearly one order of magnitude compared to geometric models $\modelz(\paramvmr,\zparamb)$ for fixed Windkessel parameters $\paramvmr$. In this section, we demonstrate that \gls*{0D} models $\modelz(\param_\text{v}, \zparamo)$ optimized with Windkessel training parameters $\param_\text{t}$ generalize well to a wide range of Windkessel validation parameters $\param_\text{v}$:
\begin{align}
\text{Compare~} \modelt(\param_\text{v}) \text{~vs.~} \modelz(\param_\text{v},\zparamo) \text{~with~} \zparamo = \arg \min_{\zparam} \Vert \vc r (\zparam, \modelt(\param_\text{t})) \Vert^2, \quad \param_\text{v} \ne \param_\text{t}.
\label{eq_0D_3D_validation}
\end{align}
This generalization is crucial to our goal of using \gls*{0D} models as a surrogate in Bayesian Windkessel calibration. If optimized \gls*{0D} models generalize well over a wide range of Windkessel parameters $\param$, we can confidently approximate the posterior distribution $\probt\approx\probz$.

Since this generalization study requires extensive sampling of computationally expensive \gls*{3D} simulation results, we select a subset of three simulation models: the aortic model of an 11-year-old human female with aortic coarctation 0104\_0001, the aortofemoral model of a 76-year-old human male with an abdominal aortic aneurysm 0140\_2001, and the pulmonary model of a 43-year-old healthy human female 0080\_0001. See Figure~\ref{fig_collage} for their geometries. These models span three different anatomies and exhibit high pressure and flow approximation errors of the geometric \gls*{0D} models of around 10\%, highlighted with arrows in Figure~\ref{fig_optimization_error}. For each model, we reused a set $\{\param_i\}$ of 50 variations of \gls*{BC} parameters from Pegolotti et al. \cite{pegolotti24}. This dataset was generated by multiplying flow rates, resistance, and capacitance values of the \glspl*{BC} parameters by a factor randomly and independently sampled from a uniform distribution between 0.8 and 1.2. In each comparison, we used identical \gls*{BC} parameters $\param_i$ in geometric \gls*{0D} models $\modelz(\param_i,\zparamb)$ and optimized ones $\modelz(\param_i,\zparamo)$ and compared them to the corresponding \gls*{3D} models $\modelt(\param_i)$. To ensure comparability, we applied the \gls*{3D} result of the first time step in the last simulated cardiac cycle as initial conditions to the \gls*{0D} simulation. The \gls*{0D} models were then evaluated for one cardiac cycle to ensure that the cardiac cycles in \gls*{0D} and \gls*{3D} were identical.

\begin{figure}[H]
\centering
\includegraphics[width=\linewidth]{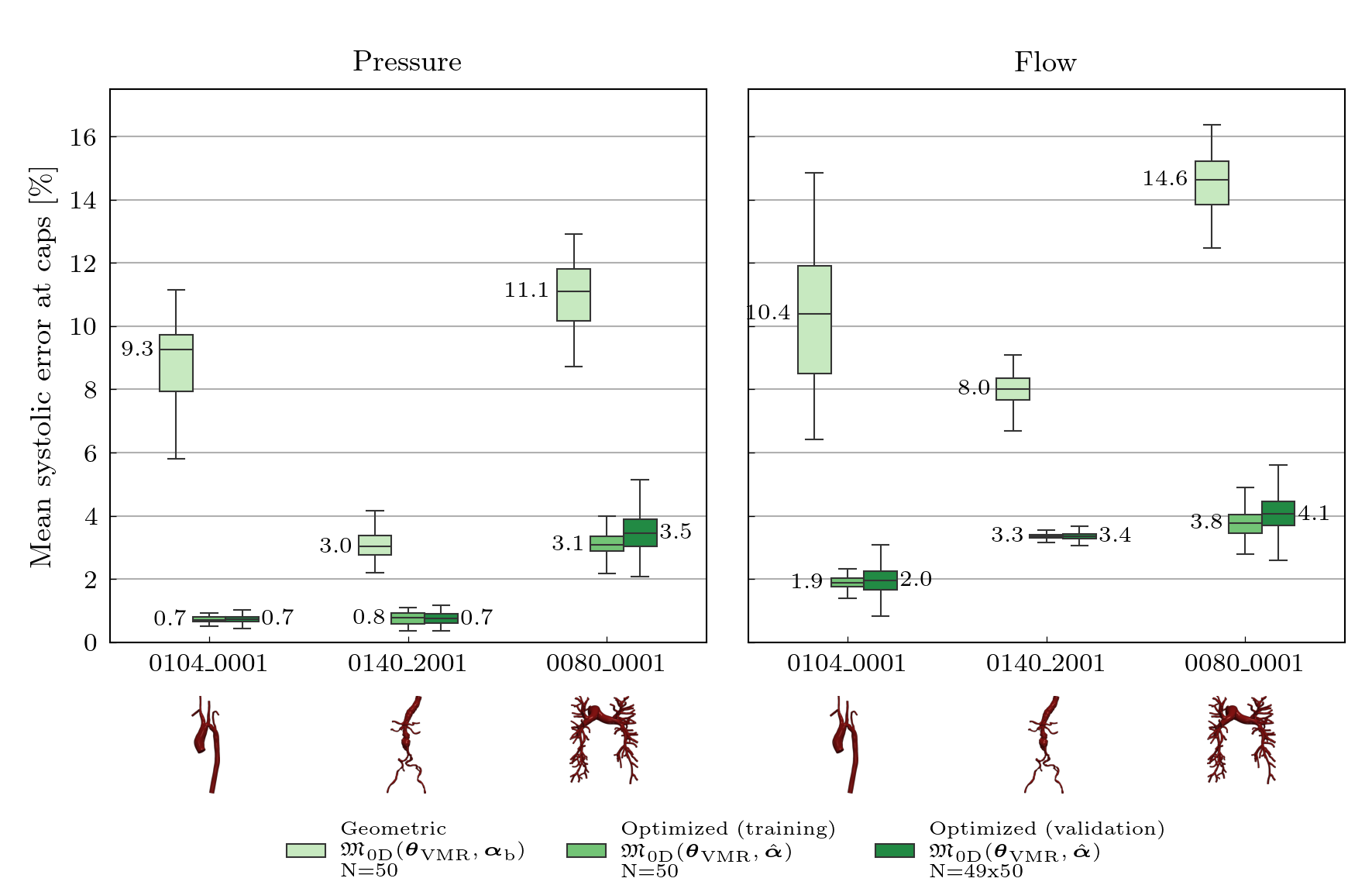}
\caption{Cross-validation of optimized \gls*{0D} model $\modelz(\paramvmr,\zparamo)$ and comparison to geometric 0D model $\modelz(\paramvmr,\zparamb)$. In each validation, the model was calibrated to the test parameters, and the result was compared to the 49 validation parameters.}
\label{fig_generalization}
\end{figure}

For each validation, we optimized the \gls*{0D} model to one \gls*{BC} training parameter set $\param_\text{t}$ and compared approximation errors for all other 49 \gls*{BC} validation parameters $\param_\text{v}$:
\begin{align}
\param_{\text{t},i} = \param_i, \quad \param_{\text{v},i} = \{\param_1,\dots,\param_{50}\} \setminus \param_i, \quad \forall i\in\{1,\dots,50\}.
\label{eq_cross_validation}
\end{align}
We perform this validation fifty times to obtain a cross-validation. Figure \ref{fig_generalization} shows the cross-validation error at systole for pressure and flow. The performance of the optimized model is further divided into performance on the training set $\{\param_{\text{t},i}\}$ and the validation set $\{\param_{\text{v},i}\}$. For reference, we also show errors for the geometric \gls*{0D} model, i.e., without optimization. There is a large variance of errors for the geometric model evaluated on all $|\{\param_i\}|=50$ \glspl*{BC} parameter sets. As observed in Section~\ref{sec_lm_training}, the error is drastically reduced on the training set of all three optimized \gls*{0D} models. Furthermore, the variance of the $|\{\param_{\text{t},i}\}| = 50$ training errors is much lower. Most importantly, the validation error is equal to or slightly larger than the training error. Despite encompassing $|\{\param_{\text{v},i}\}| = 50 \times 49 = 2450$ simulations, the variance of the validation error is only slightly larger than the one of the 50 training errors. 

We summarize two observations: First, although the optimized pulmonary models have been significantly improved, they perform worse than the aortic models. This could be related to the increased size and, thus, the complexity of the model but could also hint toward a limitation of the junction model used herein, as junctions make up a large portion of the pulmonary model domain. Second, we do not observe a notable difference in the low \gls*{0D} model error from Section~\ref{sec_lm_training} even for randomly sampled \gls*{BC} parameters. This confirms good generalization of the \gls*{LM} optimization approach, which demonstrates that we can confidently use the approximation $\modelt(\param)\approx\modelz(\param,\zparamo)$ for the Bayesian Windkessel \gls*{BC} calibration in Section~\ref{sec_bc_calibration}.



\subsection{Bayesian Windkessel calibration incorporating optimized zero-dimensional models\label{sec_bc_calibration}}
This section applies the \gls*{BC} calibration approach proposed in Section~\ref{sec_multi_fidelity} to the aortic model of an 11-year-old human female with aortic coarctation 0104\_0001. We split this section into problem-setup (\ref{sec_res_bc_setup}), results (\ref{sec_res_results}), and validation (\ref{sec_res_validation}).

\subsubsection{Problem setup \label{sec_res_bc_setup}}
Our problem setup is typical for cardiovascular fluid dynamics simulations and is often performed manually or according to well-known heuristics. Typical flow measurements are obtained from 4D Flow Magnetic Resonance Imaging at the inlet and outlets of the model \cite{nair23}. Pressure measurements at the inlet are obtained either from \textit{in vivo} catheter measurements or estimated from cuff measurements \cite{baeumler20}.

In this section, one \gls*{BC} parameter denoted as $\theta^{(i)}$ is calibrated for each outlet (i.e., each Windkessel) of the model. The parameter $\theta^{(i)}$ is the logarithmic total resistance at the respective Windkessel outlet
\begin{equation}
\theta^{(i)} = \log (R^{(i)}), \quad R^{(i)} =R_\text{p}^{(i)}+R_\text{d}^{(i)}.
\end{equation}
The logarithmic reparameterization ensures that the resistances remain positive, independent of the parameter choice, as negative resistances are unphysiological and usually lead to failed forward model evaluations. This yields the overall parameter vector
\begin{equation}
    \param = \left[\begin{matrix}
\log \left( R^{(i)}\right), & \dots, & \log \left( R^{(n_\text{out})}\right)
\end{matrix}\right]^\top,
\end{equation}
with the number of outlets $n_\text{out}$. The individual parameters of each Windkessel model, i.e., $R_\text{p}$, $R_\text{d}$, and $C$, are derived from the total resistance based on two assumptions: The ratio between distal to proximal resistance is assumed as constant and given
\begin{equation}
    \frac{R_{\text{p}, i}}{R_{\text{d}, i}} = \mathrm{const},
\end{equation}
and the time constant $\tau_i$ of each Windkessel \gls*{BC} is known and constant
\begin{equation}
    \tau_i = R_{\text{d}, i}\cdot C_i = \mathrm{const}.
\end{equation}
\reva{The respective values for $\tau_i$ and $R_{\text{p},i}/R_{\text{d},i}$ are calculated based on the existing \gls*{BC} parameters $\paramvmr$ in the Vascular Model Repository. In practice, they would be estimated from the total resistance and total capacitance of the vasculature represented by the \glspl*{BC}. However, our approach can also be used in other scenarios, e.g., simultaneously estimating capacitance $C_i$ and resistance ratio $R_{\text{p},i} / R_{\text{d},i}$ for each outlet \cite{nair24}.} The observations $\yobs$ for this example are artificially created based on one \gls*{3D} simulation for the geometry with \gls*{BC} parameters $\paramvmr$. In agreement with typical clinical measurements, the following observations have been selected: the minimum pressure at the inlet, the maximum pressure at the inlet, and the time-averaged flow rates at each outlet
\begin{equation}
    \yobs = \left[\begin{matrix}
P_{\text{in,min}}, & P_{\text{in,max}}, & Q_{\text{mean,1}}, & \dots, & Q_{\text{mean}, n_\text{outlets}}
\end{matrix}\right]^\top.
\end{equation}
The aortic anatomy 0104\_0001 has five outlets connected to Windkessel \glspl*{BC} and accordingly five calibration parameters $\theta^{(i)}$ and seven observations $y_\text{obs}^{(i)}$. It is assumed that the observations are corrupted by additive Gaussian noise as specified in Section \ref{sec_likelihood}, which is characterized by a constant signal-to-noise ratio
\begin{equation}
    \snr_i = \frac{y_{\text{obs},i}^2}{\sigma_i^2} = \text{const},
\end{equation}
for all observations $y_{\text{obs},i}$.



\subsubsection{Posterior distribution for Windkessel boundary conditions\label{sec_res_results}}
The Bayesian Windkessel calibration is performed with three signal-to-noise ratios of 100, 11.1, and 4 to investigate changes in the posterior distribution for increasing noise in observations $\yobs$. \gls*{SMC} is performed with $k=10\,000$ particles, a resampling threshold $\text{ESS}_\text{min}=5\,000$, and two rejuvenation steps per iteration. All calculations were performed on Stanford’s high performance computing cluster Sherlock using four 12-core Intel Xeon Gold 5118 CPUs. It should be noted, however, that only the \gls*{3D} simulation took advantage of all 48 available cores. We ran the \gls*{SMC} algorithm on 24 cores and the \gls*{LM} calibration on a single core. The parameter estimation for the $\snr=100$ case took a bit over 24\,h, where 1.5h was spent on the first \gls*{SMC} run (with 210\,000 \gls*{0D} evaluations), 21.5\,h on the \gls*{3D} simulation and 1.5\,h on the second \gls*{SMC} run. The \gls*{LM} calibration took only 20\,s.

\begin{figure}
\centering
\includegraphics[width=\linewidth]{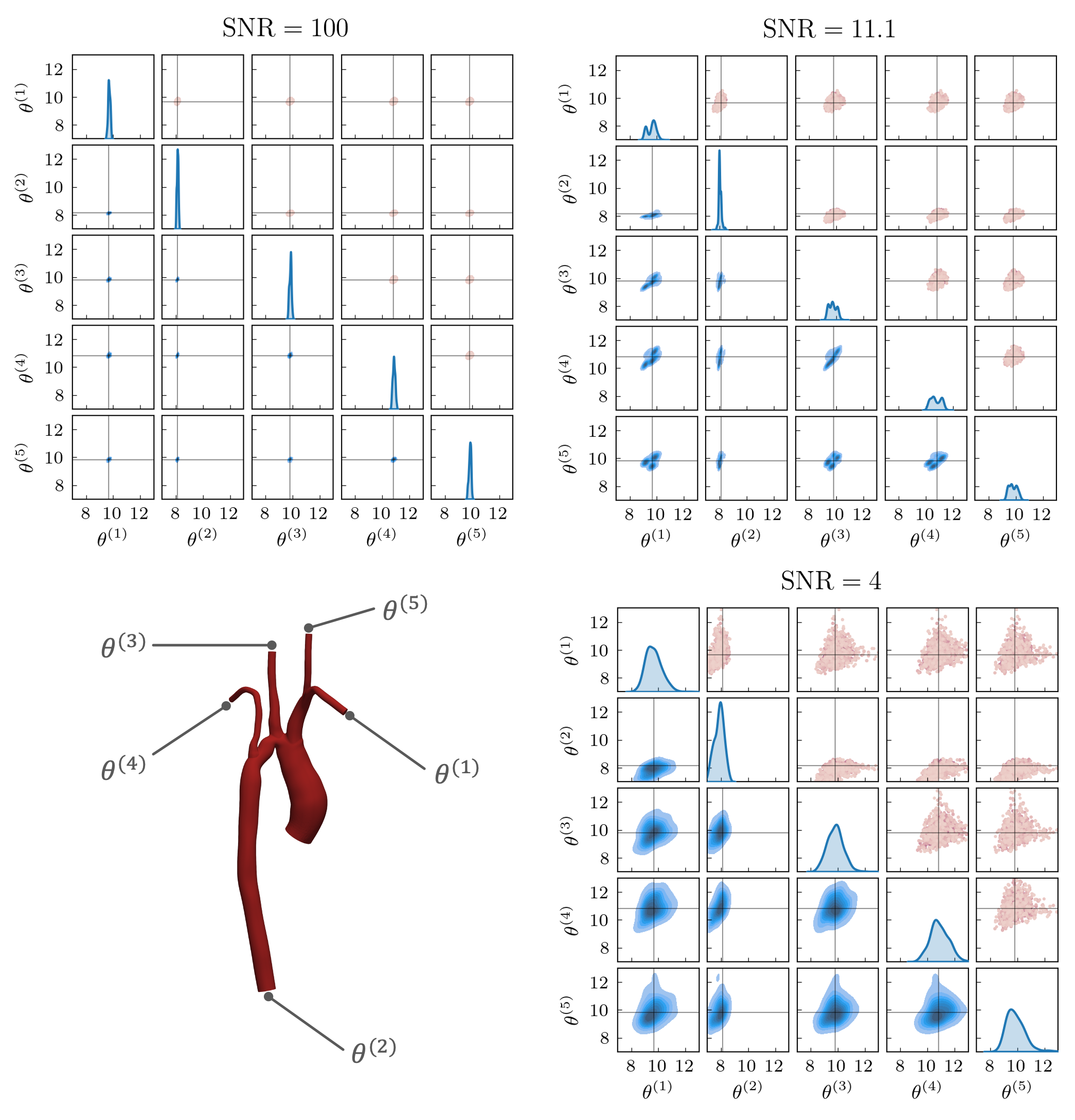}
\caption{2D marginals of the 5D posterior for all parameters $\theta^{(i)}$ and signal-to-noise ratios $\snr$ based on model 0104\_0001. The diagonal plots show the \gls*{1D} kernel density estimate. The upper right plots the particles colored by their weight, and the lower left plots the 2D  density contours. The parameter values corresponding to the ground truths are marked with crossing gray lines.}
\label{fig_posterior_0104_0001}
\end{figure}

Figure \ref{fig_posterior_0104_0001} presents the posterior densities for all parameters $\theta^{(i)}$ and signal-to-noise ratios $\snr$. All parameter distributions are concentrated around single values for the high signal-to-noise ratio $\snr=100$. The locations of the peaks match well with the parameter values used to create the artificial observations (marked by the crossing gray lines). The posterior variance is low, which matches the expectation given that the noise in observations is low. The posterior variance is generally larger for the lower single-to-noise ratio $\snr=11.1$, and some distributions exhibit multi-modality. The posterior distributions for the \gls*{BC} parameter at the descending aorta $\theta^{(2)}$ are narrower compared to the other parameters, suggesting a higher sensitivity of the model output to $\theta^{(2)}$. For the largest noise case $\snr=4$, the posterior is characterized by the largest variance.

The results show that the proposed Bayesian calibration approach successfully captured the multi-dimensional posterior distribution of the Windkessel parameters based on the optimized 0D model $\modelz(\paramvmr,\zparamo)$ and delivered robust results even for completely uninformed and broad priors. It has been demonstrated that the Bayesian calibration approach is advantageous in settings where observations are corrupted with noise. While the posterior collapses almost to a Dirac distribution in the case of a high signal-to-noise ratio in the observations, the cases with high measurement noise highlight how the uncertainty propagates to the posterior. 

Calculating the full posterior distribution provides many advantages compared to deterministic optimization strategies, where only a point estimate is obtained. First, the probabilistic calibration problem is better posed and less prone to get stuck in local optima. Second, the posterior distribution provides a quantification of confidence in the calibration result. This is instrumental for future clinical application of these models as it allows one to assess the expected reliability of a simulation model calibrated based on very sparse data. Knowing the full posterior distribution allows for more nuanced and better-informed parameter choices. Third, the posterior distributions can be used to quantify uncertainty in model predictions using forward propagation, e.g., wall shear stress in thoracic aortic aneurysms \cite{boccadifuoco18} or pressure drop over an aortic coarctation \cite{antonuccio21}. In the latter, the uncertainty in the model prediction can be taken into account while making the clinical decision of whether to intervene or not \cite{nair24}. We enabled the sampling intensive \gls*{SMC} algorithm by developing a fast \texttt{svZeroDSolver} that performed the hundreds of thousand \gls*{0D} evaluations in a bit over an hour.

\subsubsection{Comparison of 0D posterior with 3D posterior\label{sec_res_validation}}
The framework proposed in Section~\ref{sec_multi_fidelity} relies on the assumption that \gls*{0D} models $\modelz(\param,\zparamo)$ with optimized \gls*{0D} parameters $\zparamo$ approximate the \gls*{3D} model $\modelt(\param)$ accurately. This was demonstrated in a deterministic sense in Section~\ref{sec_res_0D_opti} for a wide range of Windkessel \gls*{BC} parameters $\param$. In this section, we validate that a posterior distribution based on an optimized \gls*{0D} model $\probz(\yobs|\param,\zparamo)$ agrees with the posterior distributions based on the full-dimensional \gls*{3D} model $\probt(\yobs|\param)$ given identical uncertain observations $\yobs$.

To ensure computational feasibility, we reduce the problem setup from Sections~\ref{sec_res_bc_setup} and \ref{sec_res_results} to two dimensions. The parameters $\theta^{(1)}$, $\theta^{(2)}$, $\theta^{(4)}$ and $\theta^{(5)}$ are now coupled and denoted as $\theta^{(1,2,4,5)}$ in the following and $\theta^{(3)}$ is calibrated individually. We evaluate the unnormalized posterior point-wise on a uniform 10x10 grid in the interval $[2, 8]$ for $\theta^{(1,2,4,5)}$ and $[3, 6]$ for $\theta^{(3)}$ on a geometric \gls*{0D} model $\modelz(\zparamb)$, an optimized \gls*{0D} model $\modelz(\zparamo)$, and a full-dimensional \gls*{3D} model $\modelt$. Afterward, we normalize the unnormalized posterior to yield an estimate of the true posterior distribution. A signal-to-noise ratio of $\snr=11.1\%$ is used. Performing all 100 \gls*{3D} simulations took nearly 800 hours on two 64-core AMD EPYC 7742 CPUs, while the 100 0D simulations were completed in less than a minute on a single core.

\begin{figure}
\centering
\includegraphics[width=\linewidth]{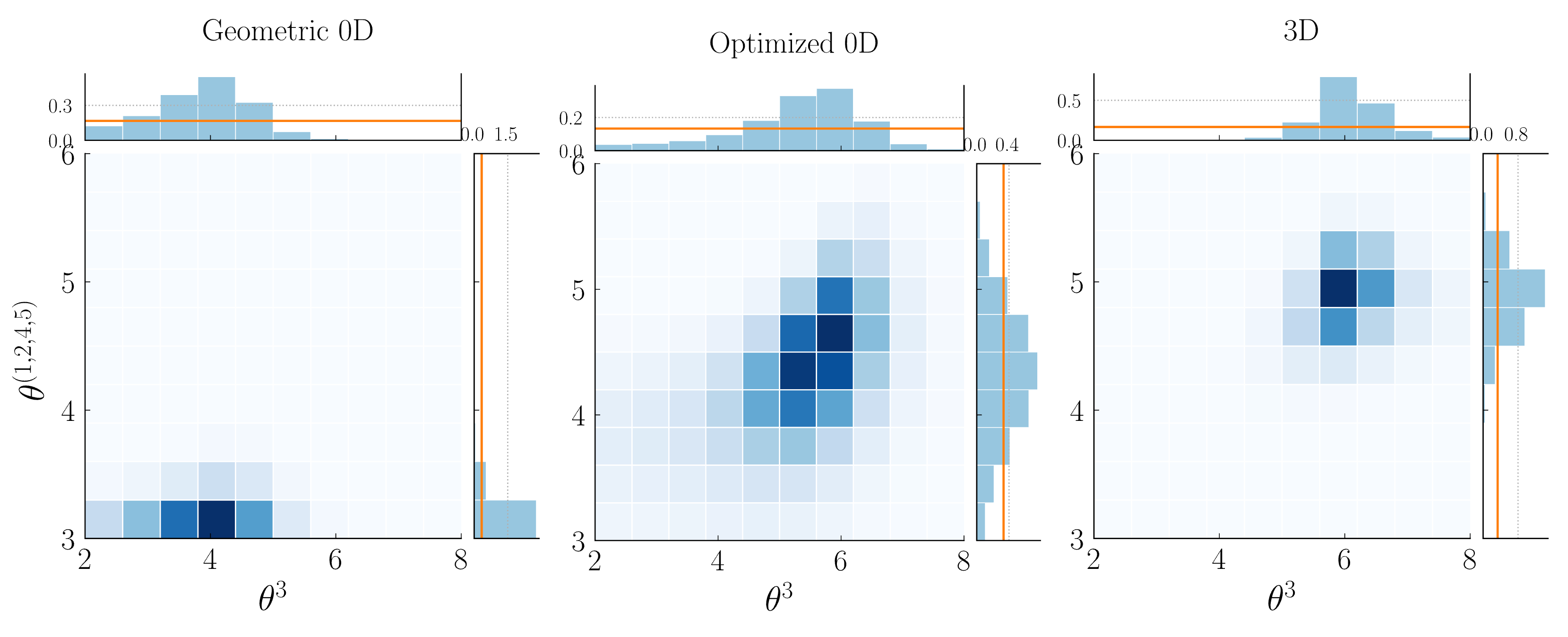}
\caption{Comparison of the posterior distribution evaluated on the geometric \gls*{0D} model, the calibrated \gls*{0D} model, and the \gls*{3D} model. The uniform prior is added in orange to the marginal distributions.}
\label{fig_grid}
\end{figure}

The resulting posterior distributions for the three models are presented in Figure~\ref{fig_grid}. The geometric and optimized \gls*{0D} models exhibit more posterior variance than the \gls*{3D} model. While the geometric \gls*{0D} model posterior shows no overlap with the \gls*{3D} posterior, the optimized \gls*{0D} model has a much-improved agreement with the latter. Despite the slight elongation of the posterior for the \gls*{0D} models, \gls*{3D} and optimized \gls*{0D} models attribute most of the probability mass in a similar region close to the coordinate (6, 5). As seen in Section \ref{sec_lm_training}, model 0104\_0001 was intentionally selected as one of the worst-performing \gls*{0D} models. However, it still achieved a good approximation of the \gls*{3D} posterior, especially compared to the geometric \gls*{0D} model, with a tendency to overestimate the uncertainty in the calibration result.

The optimized \gls*{0D} posterior deviates slightly from the high-fidelity \gls*{3D} posterior but clearly encapsulates the latter. In contrast, the geometric \gls*{0D} model completely fails to provide a viable surrogate for the \gls*{3D} posterior. While the geometric \gls*{0D} model attributes the majority of its probability density to wrong parameter choices and, thus, provides confidence in incorrect results, the optimized 0D model merely overpredicts the posterior variance. \revb{This discrepancy is due to the residual \gls*{0D}-\gls*{3D} approximation error after optimization.} Informing the \gls*{0D} model by a single \gls*{3D} simulation result was sufficient to increase the agreement between \gls*{0D} model and \gls*{3D} model to the extent that the \gls*{0D} model can confidently be used for parameter calibration without requiring further \gls*{3D} evaluations. These results confirm that the optimized \gls*{0D} models are not only superior to the geometric \gls*{0D} models for approximating the \gls*{3D} posterior but are essential. Moreover, it has been confirmed that the approach proposed in this work allows for accurate and efficient probabilistic parameter calibration for Windkessel \glspl*{BC}.

\section{Limitations and outlook\label{sec_conclusion}}


We demonstrated that a \gls*{LM} optimization approach can derive highly accurate \gls*{0D} models based on a \gls*{3D} simulation of a single heartbeat. The optimized \gls*{0D} models showed great agreement with the physiological behavior of \gls*{3D} models even during substantial changes in the \glspl*{BC}. The optimization also proved effective for automatically determining parameters for \gls*{0D} junctions, which is a known challenge for cardiovascular reduced-order models. The results provide valuable insights regarding the potential to improve the automatic creation of \gls*{LPN} models. This could be used to develop more accurate 0D element parameters that are informed by both geometry and hemodynamics or as training data for machine-learning-based approaches to \gls*{0D} model generation \cite{rubio24}. 

Despite the significant reduction in \gls*{0D}-\gls*{3D} approximation errors, there are still some deviations between the optimized \gls*{0D} models and the \gls*{3D} models, especially for pulmonary models. While it is not possible to achieve perfect agreement between \gls*{3D} and \gls*{0D} models, this indicates that improvements to the \gls*{0D} models themselves may need to be investigated to include more nonlinear terms that better capture \gls*{3D} flow physics. A structured analysis of the physiological effects currently not accounted for by the \gls*{0D} elements could result in better agreement. Additionally, a probabilistic optimization approach could provide more information about the identifiability of the individual 0D parameters based on the \gls*{3D} training data. The \gls*{0D} model optimization has been performed using one cardiac cycle of a periodically converged simulation. Future studies should investigate whether the amount of \gls*{3D} data necessary to sufficiently inform the \gls*{0D} model optimization can be further reduced, i.e., a few \gls*{3D} time steps may suffice. This would reduce the computational expense of the algorithm even further.

We demonstrated that optimized \gls*{0D} models can be deployed for accurate and efficient Bayesian Windkessel calibration using \gls*{SMC}. \reva{While the optimized 0D posterior slightly overestimates uncertainty compared to the ground-truth 3D posterior, it significantly adds more nuance than the state-of-the-art deterministic calibration.} The Bayesian approach not only provides more robustness for the calibration, but the resulting posterior distributions also provide a powerful tool for making more informed parameter decisions in patient-specific cardiovascular modeling. While deterministic calibration results can only be interpreted in one way (as the final parameter choice), the posterior distributions can be processed to retrieve the optimal parameter choice in view of multiple aspects (e.g. risk-assessment, feasibility analysis). The posterior also provides additional information about the uncertainty in the parameters given potentially noisy clinical measurements. This is especially relevant in a clinical setting where data is usually sparse and noisy. The results can also be used to decide whether a parameter choice can be made confidently based on the given clinical measurement or if more data is required.

We evaluated the proposed calibration framework on artificially created measurement data. While the problem setup aimed to match a common calibration scenario, further work could focus on using subject data. Applying the proposed calibration framework to different setups can validate its robustness on realistic clinical data. Moreover, we limited our investigations to rigid-wall \gls*{3D} models of the vasculature. While this is a valid assumption for the stiffer vasculature or lower pressures, it is not generally applicable. \reva{Nevertheless, despite wall compliance being near zero, we demonstrated that our \gls*{0D} parameter optimization can robustly identify vessel capacity. Further, we demonstrated that our forward and inverse \gls*{0D} models can incorporate vessel wall elasticity and highlighted differences between rigid and compliant wall pressure curves in a previous publication \cite{pfaller22}. We thus expect that our calibration framework will be applicable to compliant-wall simulations as well.}

The posterior distribution derived based on the optimized 0D models has been compared against the high-fidelity \gls*{3D} posterior and proved to be a valid surrogate for the latter. The poor performance of the geometric 0D model highlighted that the 0D-3D optimization is essential for performing Bayesian calibration on 0D models. \reva{Our framework can be improved by incorporating more than one high-fidelity 3D simulation in clinical scenarios that require a more accurate calibration.} \revb{A multi-fidelity approach could model the 0D-3D discrepancy to generate a more accurate surrogate for the 3D model \cite{lee23,biehler2015towards,seo20,fleeter20}.} Alternatively, further work could generalize the algorithm presented herein using an expectation-maximization approach that treats the 0D model parameters $\zparam$ as latent variables that are consistently optimized as part of the inverse problem. \reva{This algorithm has guaranteed convergence properties under certain conditions and would allow for a more robust Bayesian \gls*{0D} parameter optimization.} \revb{This full Bayesian treatment also allows to quantify the uncertainty of the 0D parameter optimization within the \gls*{BC} calibration.}


\section*{Acknowledgements}
This work was supported by grants K99HL161313, R01HL167516, and R01HL141712 from the National Heart, Lung, and Blood Institute and the Stanford Maternal and Child Health Research Institute. DES acknowledges partial support from NSF CAREER award \#1942662 and NSF CDS\&E award \#2104831. WAW and JN acknowledge support from the European Research Council (grant agreement No. 101021526-BREATHE). The authors gratefully acknowledge the Stanford Research Computing Center and the San Diego Super Computing Center (SDSC) Expanse cluster for providing the computational resources necessary for the numerical simulations presented in this work.

\newpage
\appendix
\glsresetall
\section{Overview of cardiovascular models}
\label{sec_models}
\begin{figure}[H]
\centering
\includegraphics[width=.93\linewidth]{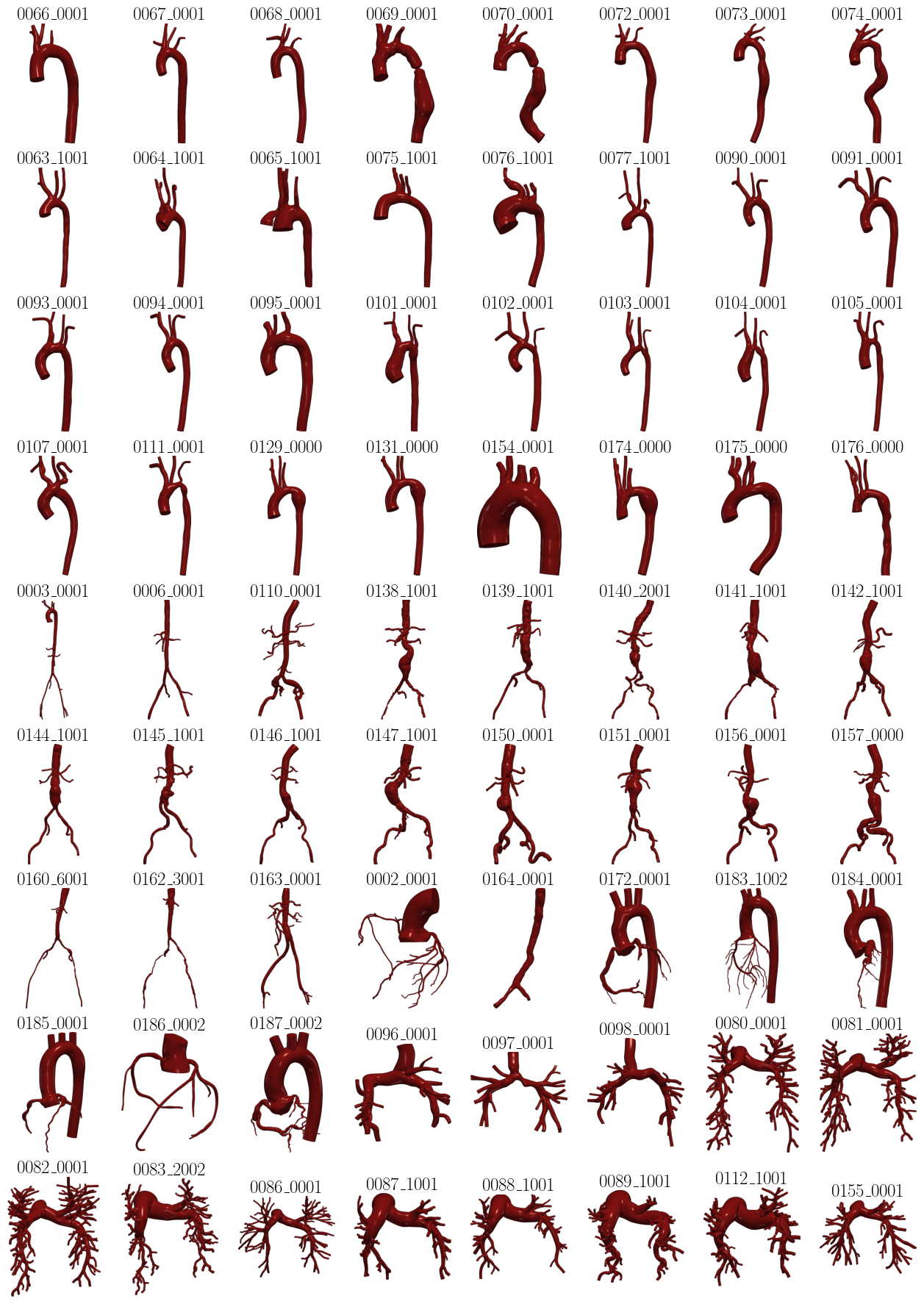}
\caption{Overview of models used in this work, sorted by anatomy. This set is identical to the one in Pfaller et al. \cite{pfaller22}.\label{fig_collage}}
\end{figure}

Figure~\ref{fig_collage} shows an overview of all 72 models used in this work, identical to the ones in Pfaller et al. \cite{pfaller22}. They are openly available from the Vascular Model Repository \cite{vmr}.

\section{Computational details of zero-dimensional models\label{sec_matrices}}
This section contains the \gls*{0D} time integration scheme (\ref{sec_gen_alpha}) and the system matrices for the \texttt{BloodVessel} (\ref{sec_0D_ele_vessel}), \texttt{BloodVesselJunction} (\ref{sec_0D_ele_junction}), and Windkessel (\ref{sec_0D_ele_windkessel}) lumped-parameter network elements as implemented in \texttt{svZeroDSolver} \cite{svzerodsolver}.

\subsection{Forward time integration\label{sec_gen_alpha}}
We use a generalized-$\alpha$ time integration \cite{chung93,jansen00} and solve for the solution vector $\vc y$ in each time step $n$ using the Newton-Raphson method,
\begin{align}
\text{Predict:~}\quad
\dot{\vc y}_{n+1}^0 &= \frac{\gamma-1}{\gamma} \dot{\vc y}_n,\\
\vc y_{n+1}^0 &= \vc y_n,\\[.5cm]
\text{Initiate:~}\quad
\dot{\vc y}_{n+\alpha_m}^i &= \dot{\vc y}_n + \alpha_m \left( \dot{\vc y}_{n+1}^i - \dot{\vc y}_n  \right),\\
\vc y_{n+\alpha_f}^i &= \vc y_n + \alpha_f \left( \vc y_{n+1}^i - \vc y_{t} \right),\\[.5cm]
\text{Solve:~}\quad
\vc K(\vc y_{n+\alpha_f}^i, \dot{\vc y}_{n+\alpha_m}^i) \cdot \Delta \dot{\vc y}_{n+1}^i &= - \vc r(\vc y_{n+\alpha_f}^i, \dot{\vc y}_{n+\alpha_m}^i),\\[.5cm]
\text{Update:~}\quad
\dot{\vc y}_{n+1}^{i+1} &= \dot{\vc y}_{n+1}^i + \Delta \dot{\vc y}_{n+1}^i,\\
\vc y_{n+1}^{i+1} &= \vc y_{n+1}^i + \Delta \dot{\vc y}_{n+1}^i \, \gamma \Delta t_n,
\end{align}
with time step $\Delta t_n$, residual $\vc r$, and Jacobian $\vc K$. We derive the generalized-$\alpha$ parameters $\alpha_f$, $\alpha_m$, and $\gamma$ from the spectral radius $\rho_\infty$
\begin{align}
\alpha_m = \frac{1 - \rho_\infty}{2(1 + \rho_\infty)}, \quad \alpha_f = \frac{1}{1+\rho_\infty}, \quad \gamma &= \frac{1}{2} + \alpha_m - \alpha_f.
\end{align}

\subsection{Inverse parameter optimization \label{sec_optimization}}
We use the \gls*{LM} algorithm \cite{levenberg44,marquardt63} to solve the least squares problem \eqref{eq_0D_least_squares} iteratively for the optimal parameter set $\zparamo$,
\begin{align}
\text{update~ } \zparam^{i+1} &= \zparam^{i}+\Delta \zparam^{i+1},\\
\text{solve~ }
\left[
\vc J^{\mathrm{T}} \vc J + \lambda \operatorname{diag} 
\left(\vc J^{\mathrm{T}} \vc J\right) 
\right]^{i}
\cdot
\Delta \zparam^{i+1} 
&= - \left[\vc J^{\mathrm{T}}
\vc r\right]^{i}, \quad \lambda^{i}=\lambda^{i-1}
\cdot\left\|\left[\vc J^{\mathrm{T}} \vc r\right]^{i}\right\|_2
/\left\|\left[\vc J^{\mathrm{T}} \vc r\right]^{i-1}\right\|_2.
\label{eq_lm}
\end{align}
The \gls*{LM} algorithm approximates the Hessian from the Jacobian matrix $\vc J$ as $\Delta S \approx \vc J^{\mathrm{T}} \vc J$. The second term introduces damping with factor $\lambda$, which we update based on the change in gradient $\nabla S = \vc J^{\mathrm{T}} \vc r$. Thus, the damping parameter approaches $\lambda\to0$ as the parameter set $\zparam\to\zparamo$ approaches the optimal solution. For $\lambda\to\infty$, we obtain the steepest descent method, and for $\lambda = 0$, we obtain the Gauss-Newton method. We terminate the optimization \eqref{eq_lm} when both $L^2$ norms,
\begin{align}
\left\|\left[\vc J^{\mathrm{T}} \vc r\right]^{\mathrm{i}}\right\|_2
< \operatorname{tol}_{\text{grad}}^{\zparam} 
\text { and }\left\|\Delta \zparam^{\mathrm{i}+1}\right\|_2
< \operatorname{tol}_{\text{inc}}^{\zparam},
\end{align}
are below the tolerances for gradient and increment, $\operatorname{tol}_{\text{grad}}^{\zparam}$ and $\operatorname{tol}_{\text{inc}}^{\zparam}$, respectively. In this work, we chose the \gls*{LM} optimization parameters of initial damping factor $\lambda^0=1$, gradient tolerance $\operatorname{tol}_{\text{grad}}^{\zparam} = 10^{-5}$, increment tolerance $\operatorname{tol}_{\text{inc}}^{\zparam} = 10^{-10}$, and a maximum number of iterations of $100$.

\subsection{BloodVessel element\label{sec_0D_ele_vessel}}
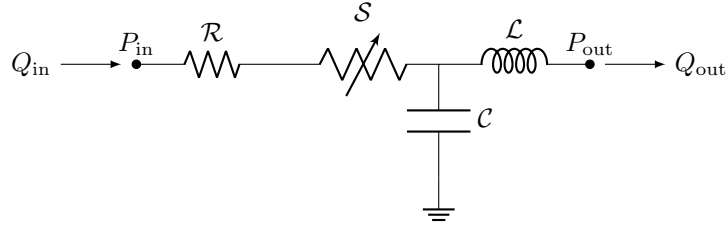
\begin{figure}[H]
\centering
\begin{circuitikz}
\draw node[left] {$Q_\text{in}$} [-latex] (0,0) -- (0.8,0);
\draw (1,0) node[anchor=south]{$P_\text{in}$}
to [R, l=$\od{R}$, *-] (3,0)
to [vR, l=$\od{S}$, -] (5,0) 
(5,0) to [L, l=$\od{L}$, -*] (7,0)
node[anchor=south]{$P_\text{out}$}
(5,0) to [C, l=$\od{C}$, -] (5,-1.5)
node[ground]{};
\draw [-latex] (7.2,0) -- (8,0) node[right] {$Q_\text{out}$};
\end{circuitikz}
\caption{The \texttt{BloodVessel} element with resistance $\od{R}$, stenosis $\od{S}$, inertance $\od{L}$, and capacitance $\od{C}$.\label{fig_0D_vessel}}
\end{figure}

We model the blood vessel branches with the \texttt{BloodVessel} element, whose circuit is visualized in Figure~\ref{fig_0D_vessel}.
The governing equations are given by
\begin{align}
- (\od{R} + \od{S} |Q_\text{in}|) \, Q_\text{in} - \od{L} \, \dot{Q}_\text{out} - \Delta P &= 0,\\
\od{C} (\od{R} + 2 \od{S} |Q_\text{in}|) \, \dot{Q}_\text{in} - \od{C} \dot{P}_\text{in} - \Delta Q &= 0.
\label{eq_bloodvessel}
\end{align}
The remainder of this section contains the forward (\ref{sec_0D_ele_vessel_forward}) and inverse (\ref{sec_0D_ele_vessel_inverse}) problem of the \texttt{BloodVessel} element.

\subsubsection{Forward problem\label{sec_0D_ele_vessel_forward}}
The vector of element unknowns is
\begin{align}
\vc{y}^{(e)} &= 
\left[\begin{array}{llll}
P_\text{in}^{(e)} & Q_\text{in}^{(e)} &
P_\text{out}^{(e)} & Q_\text{out}^{(e)}
\end{array}\right]^\mathrm{T}
\end{align}
The \texttt{BloodVessel} element matrices for the forward problem~\eqref{eq_0D_forward} are
\begin{align}
\vc{E}^{(e)} &= 
\left[\begin{array}{cccc}
0 & 0 & 0 & -\od{L} \\
-\od{C} & \od{C}\od{R} & 0 & 0
\end{array}\right]^{(e)}, \\
\vc{F}^{(e)} &= 
\left[\begin{array}{cccc}
1 & - \od{R} & -1 & 0 \\
0 & 1 & 0 & -1
\end{array}\right]^{(e)}, \\
\vc{c}^{(e)} &= \od{S} \, |Q_\text{in}|
\left[\begin{array}{c}
-Q_\text{in}\\
2\od{C}\dot{Q}_\text{in}
\end{array}\right]^{(e)}.
\end{align}
Only the stenosis term $\od{S}$ contributes to the nonlinear term $\vc c$. Its linearizations are
\begin{align}
\left( \pfrac{\vc c}{\dot{\vc y}} \right)^{(e)} &= \od{S} \,|Q_\text{in}|
\left[\begin{array}{cccc}
0 & 0 & 0 & 0 \\
0 & 2\od{C} & 0 & 0
\end{array}\right]^{(e)},\\
\left( \pfrac{\vc c}{\vc y} \right)^{(e)} &= \od{S} \, \text{sgn} (Q_\text{in})
\left[\begin{array}{cccc}
0 & - 2Q_\text{in} & 0 & 0 \\
0 & 2\od{C}\dot{Q}_\text{in} & 0 & 0
\end{array}\right]^{(e)}.
\end{align}

\subsubsection{Inverse problem\label{sec_0D_ele_vessel_inverse}}
The vector of element unknowns is
\begin{align}
\zparam^{(e)} &= 
\left[\begin{array}{llll}
\od{R}^{(e)} & \od{C}^{(e)} & \od{L}^{(e)} & \od{S}^{(e)}
\end{array}\right]^\mathrm{T},
\end{align}
containing flow and pressure for the inlet and $n$ outlets. The \texttt{BloodVessel} element Jacobian matrix for the inverse problem~\eqref{eq_0D_inverse} is
\begin{align}
\vc{J}^{(e)} = 
\left[\begin{array}{cccc}
-Q_\text{in} & 0 & -\dot{Q}_\text{out} & -|Q_\text{in}|Q_\text{in} \\
\od{C} \, \dot{Q}_\text{in} & -\dot{P}_\text{in} + (\od{R} + 2 \od{S} \, |Q_\text{in}|) \, \dot{Q}_\text{in} & 0 & 2\od{C}|Q_\text{in}|\dot{Q}_\text{in}
\end{array}\right]^{(e)}.
\end{align}

\subsection{BloodVesselJunction element\label{sec_0D_ele_junction}}
\begin{figure}[H]
\centering
\begin{circuitikz}
\draw node[left] {$Q_\text{in}$} [-latex] (0,0) -- (0.8,0);
\draw (1,0.1) node[anchor=south]{$P_\text{in}$};
\draw (1,0) to [short, *-] (2.5,0.75);
\draw (1,0) to [short, *-] (2.5,-0.75);
\draw (2.5,0.75) node[anchor=south]{} to [generic, l^=$BV_{1}$, -*] (4.5,0.75); 
\draw (2.4,0.75) node[anchor=south]{};
\draw (4.6,0.75) node[anchor=south] {$P_{\text{out},1}$};
\draw (2.5,-0.75) node[anchor=south]{} to [generic, l_=$BV_{n}$, -*] (4.5,-0.75);
\draw (2.4,-0.75) node[anchor=north]{};
\draw (4.6,-0.75) node[anchor=north] {$P_{\text{out},n}$};
\draw [-latex] (4.7,0.75) -- (5.5,0.75) node[right] {$Q_{\text{out},1}$};
\draw [-latex] (4.7,-0.75) -- (5.5,-0.75) node[right] {$Q_{\text{out},n}$};
\draw (3.5,-0.1) node[anchor=south]{\dots};
\end{circuitikz}\\
\begin{circuitikz} \draw
node[left] {$Q_\text{in}$} [-latex] (0,0) -- (0.8,0);
\draw (1,0) node[anchor=south]{$P_\text{in}$}
to [R, l=$\od{R}_i$, *-] (3,0)
to [vR, l=$\od{S}_i$, -] (5,0)
(5,0) to [L, l=$\od{L}_i$, -*] (7,0)
node[anchor=south]{$P_{\text{out},i}$};
\draw [-latex] (7.2,0) -- (8,0) node[right] {$Q_{\text{out},i}$};
\end{circuitikz}
\caption{The \texttt{BloodVesselJunction} element (top), connecting an inlet to $n$ outlets. An individual inlet-outlet connection $BV_i$ is shown at the bottom, with resistance $\od{R}_i$, stenosis $\od{S}_i$, and inertance $\od{L}_i$.\label{fig_0D_junction}}
\end{figure}
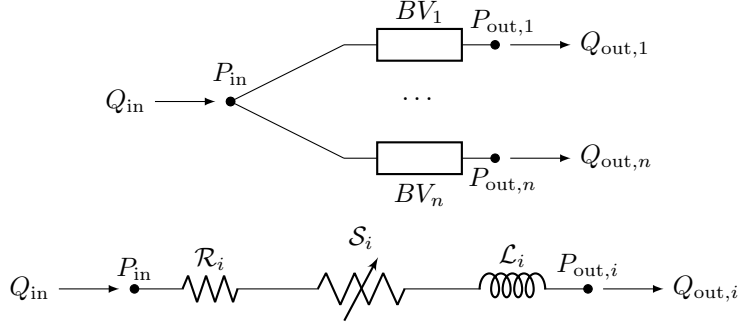

We model the junctions between blood vessel branches with the \texttt{BloodVesselJunction} element. It is visualized in Figure~\ref{fig_0D_junction} (top) for an inlet connecting to $1,\dots,n$ outlets. An individual junction model connecting the inlet to outlet~$i$ is visualized in Figure~\ref{fig_0D_junction} (bottom). 
The governing equations for the \texttt{BloodVesselJunction} are
\begin{align}
Q_\text{in}-\sum_i^n Q_{\text{out}, i} &= 0,\label{eq_bloodvesseljunction_mass}\\
\Delta P_i  + (\od{R} + \od{S}|Q_{\text{out},i}|) \, Q_{\text{out},i} + \od{L} \, \dot{Q}_{\text{out},i} &= 0,  \quad \forall \, i = 1, \dots,n.
\label{eq_bloodvesseljunction}
\end{align}
The remainder of this section contains the forward (\ref{sec_0D_ele_junction_forward}) and inverse (\ref{sec_0D_ele_junction_inverse}) problem of the \texttt{BloodVesselJunction} element.

\subsubsection{Forward problem}
\label{sec_0D_ele_junction_forward}
The vector of element unknowns is
\begin{align}
\vc{y}^{(e)} &= 
\left[\begin{array}{lllllll}
P_\text{in}^{(e)} & Q_\text{in}^{(e)} &
P_{\text{out},1}^{(e)} & Q_{\text{out},1}^{(e)} &
P_{\text{out},2}^{(e)} & Q_{\text{out},2}^{(e)} & \dots
\end{array}\right]^\mathrm{T}.
\end{align}
The \texttt{BloodVesselJunction} element matrices for the forward problem~\eqref{eq_0D_forward} are
\begin{align}
\vc{E}^{(e)} &= 
\left[\begin{array}{ccccccc}
0 & 0 & 0 & 0 & 0 & 0 & \dots \\
0 & 0 & 0 & -\od{L}_1 & 0 & 0 & \\
0 & 0 & 0 & 0 & 0 & -\od{L}_2 & \\
\vdots & & & & & & \ddots 
\end{array}\right]^{(e)}, \\
\vc{F}^{(e)} &= 
\left[\begin{array}{ccccccc}
0 & 1 & 0 & -1 & 0 & -1 & \dots \\
1 & 0 & -1 & -\od{R}_1 & 0 & 0 & \\
1 & 0 & 0 & 0 & -1 & -\od{R}_2 & \\
\vdots & & & & & & \ddots
\end{array}\right]^{(e)}, \\
\vc{c}^{(e)} &= 
\left[\begin{array}{c}
0 \\
- \od{S}_1 |Q_{\text{in},1}| \, Q_{\text{in},1} \\
- \od{S}_2 |Q_{\text{in},2}| \, Q_{\text{in},2} \\
\vdots
\end{array}\right]^{(e)},
\end{align}
with the first rows in each component containing the conservation of mass equation \eqref{eq_bloodvesseljunction_mass}. The linearization of the nonlinear term $\vc c$ is
\begin{align}
\left( \pfrac{\vc c}{\dot{\vc y}} \right)^{(e)} &= \vc 0,\\
\left( \pfrac{\vc c}{\vc y} \right)^{(e)} &= 2
\left[\begin{array}{cccccccc}
0 & 0 & 0 & 0 & 0 & 0 & \dots\\
0 & 0 & 0 & - \od{S}_1 |Q_{\text{in},1}| & 0 & 0 & \\
0 & 0 & 0 & 0 & 0 & - \od{S}_1 |Q_{\text{in},2}| & \\
\vdots & & & & & & \ddots \\
\end{array}\right]^{(e)}.
\end{align}

\subsubsection{Inverse problem}
\label{sec_0D_ele_junction_inverse}
The vector of element unknowns is
\begin{align}
\zparam^{(e)} &= 
\left[\begin{array}{lllllllll}
\od{R}_1^{(e)} & \od{R}_2^{(e)} & \dots &
\od{L}_1^{(e)} & \od{L}_2^{(e)} & \dots & 
\od{S}_1^{(e)} & \od{S}_2^{(e)} & \dots
\end{array}\right]^\mathrm{T},
\end{align}
consisting of $3n$ unknowns for a junction with $n$ outlets. The \texttt{BloodVesselJunction} element Jacobian matrix for the inverse problem~\eqref{eq_0D_inverse} is
\begin{align}
\vc{J}^{(e)} = 
\left[\begin{array}{ccccccccc}
0 & 0 & \dots & 0 & 0 & \dots & 0 & 0 & \dots \\
-Q_{\text{out},1} & 0 & & -\dot{Q}_{\text{out},1} & 0 & & |Q_{\text{in},1}| \, Q_{\text{in},1} & 0 & \\
0 & -Q_{\text{out},2} & & 0 & -\dot{Q}_{\text{out},2} & & 0 & |Q_{\text{in},2}| \, Q_{\text{in},2} & \\
\vdots & & \ddots & \vdots & & \ddots & \vdots & & \ddots
\end{array}\right]^{(e)}.
\end{align}

\subsection{Windkessel element\label{sec_0D_ele_windkessel}}
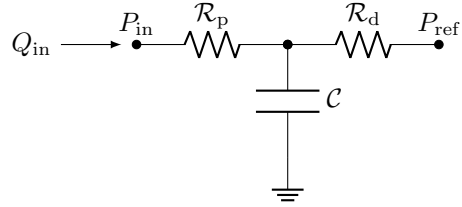
\begin{figure}[H]
\centering
\begin{circuitikz}
\draw node[left] {\quad $Q_{\text{in}}$} [-latex] (0,0) -- (0.8,0);
\draw (1,0) node[anchor=south]{$P_{\text{in}}$}
to [R, l=$\od{R}_\text{p}$, *-] (3,0)
to [R, l=$\od{R}_\text{d}$, *-*] (5,0)
node[anchor=south]{$P_{\text{ref}}$}
(3,0) to [C, l=$\od{C}$, *-] (3,-1.5)
node[ground]{};
\end{circuitikz}
\caption{Three-element \texttt{Windkessel} \gls*{BC} \label{fig_0D_windkessel}}
\end{figure}

We use a three-element Windkessel visualized in Figure~\ref{fig_0D_windkessel}.
It is governed by the equation
\begin{align}
(\od{R}_\text{p} + \od{R}_\text{d}) \, Q_\text{in} - P_\text{in} + P_\text{ref} - \od{R}_\text{d} \, \od{C} \, \dot{P}_\text{in} + \od{R}_\text{p} \od{R}_\text{d} \, \dot{Q}_\text{in} &= 0.
\label{eq_windkessel}
\end{align}
consisting of a capacitance $\od{C}$, and proximal and distal resistances $\od{R}_\text{p}$ and $\od{R}_\text{d}$, respectively \cite{vignonclementel06}. In our examples, the reference pressure is $P_\text{ref}=0$ at all outlets. The vector of element unknowns for the Windkessel \gls*{BC} is
\begin{align}
\vc{y}^{(e)} &= 
\left[\begin{array}{ll}
P_\text{in}^{(e)} & Q_\text{in}^{(e)}
\end{array}\right]^\mathrm{T}.
\end{align}
The forward problem~\eqref{eq_0D_forward} is given by
\begin{align}
\vc{E}^{(e)} &= 
\left[\begin{array}{cc}
-\od{R}_\text{d} \, \od{C} & \od{R}_\text{p} \, \od{R}_\text{d}
\end{array}\right]^{(e)}, \\
\vc{F}^{(e)} &= 
\left[\begin{array}{cccc}
-1 & \od{R}_\text{p} + \od{R}_\text{d}
\end{array}\right]^{(e)}, \\
\vc{c}^{(e)} &=
\left[\begin{array}{cccc}
P_\text{ref}
\end{array}\right]^{(e)}.
\end{align}
The derivatives of the nonlinear term $\vc c$ vanish:
\begin{align}
\left( \pfrac{\vc c}{\dot{\vc y}} \right)^{(e)} =
\left( \pfrac{\vc c}{\vc y} \right)^{(e)} = \vc 0.
\end{align}

\section{Sequential Monte Carlo algorithm\label{sec_smc_details}}
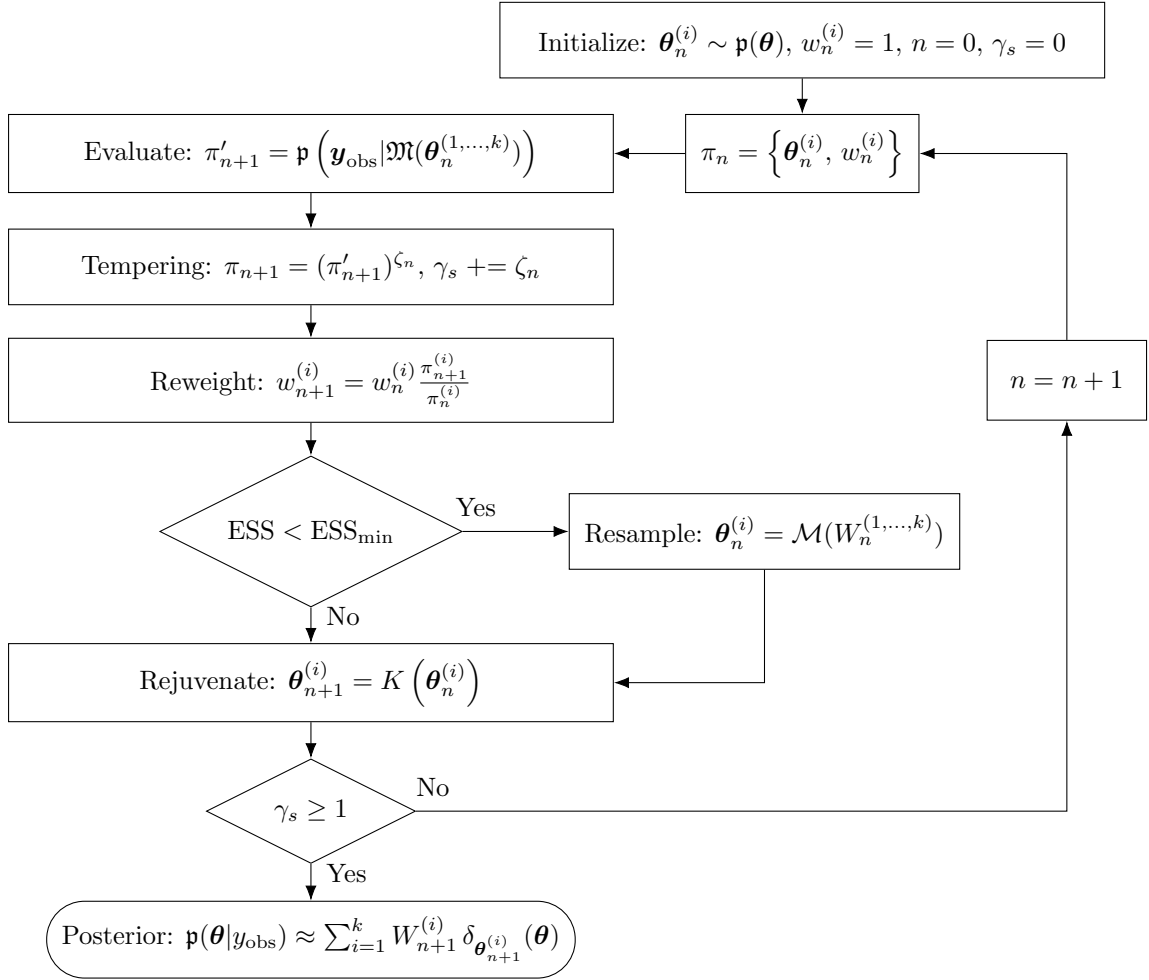
\begin{figure}
    \centering
    \begin{tikzpicture}[auto, >={Latex[length=2mm]}]
        \tikzset{every node/.style={inner sep=0.2cm}}
        \node [block, minimum height=1cm, minimum width=8cm] (step_1) {Initialize: $\param_n^{(i)}\sim \prob(\param), \, w_n^{(i)}=1, \, n=0, \,\gamma_s = 0$};
        \node [block, below of=step_1, minimum height=1cm, node distance=1.5cm] (step_2) {$\pi_n = \left\{\param_n^{(i)}, \, w_n^{(i)}\right\}$};
        \node [block, left of=step_2, minimum height=1cm, minimum width=8cm, node distance=6.5cm] (step_3) {Evaluate: $\pi_{n+1}^\prime = \prob\left(\yobs|\model(\param_{n}^{(1,\dots,k)})\right)$};
        \node [block, below of=step_3, minimum height=1cm, minimum width=8cm, node distance=1.5cm] (step_4) {Tempering: $\pi_{n+1} = (\pi_{n+1}^\prime)^{\zeta_n}, \, \gamma_s \pluseq \zeta_n$};
        \node [block, below of=step_4, minimum height=1cm, minimum width=8cm, node distance=1.5cm] (step_5) {Reweight: $w_{n+1}^{(i)}=w_n^{(i)} \frac{\pi_{n+1}^{(i)}}{\pi_{n}^{(i)}}$};
        \node [draw, diamond, aspect=2, below of=step_5, minimum height=1.8cm, minimum width=4.0cm, node distance=2.0cm] (decision_1) {$\mathrm{ESS}<\mathrm{ESS_{\text{min}}}$};
        \node [block, right of=decision_1, minimum height=1cm, node distance=6cm] (step_6) {Resample: $\param_{n}^{(i)}=\mathcal{M}(W_{n}^{(1,\dots,k)})$};
        \node [block, below of=decision_1, minimum height=1cm, minimum width=8cm, node distance=2.0cm] (step_7) {Rejuvenate: $\param_{n+1}^{(i)}=K\left(\param_{n}^{(i)}\right)$};
        \node [draw, diamond, aspect=2, below of=step_7, minimum height=1.3cm, minimum width=2.5cm, node distance=1.7cm] (decision_2) {$\gamma_s \geq 1$};
        \node [block, rounded corners=0.5cm, below of=decision_2, minimum height=1cm, node distance=1.7cm] (end) {Posterior: $\prob(\param|y_{\text{obs}}) \approx \sum_{i=1}^k W_{n+1}^{(i)} \, \delta_{\param_{n+1}^{(i)}}(\param)
        $};
        \node [block, right of=step_5, node distance=10cm] (step_8) {$n=n+1$};
    
        \draw [draw,->] (step_1) -- node {} (step_2);
        \draw [draw,->] (step_2) -- node {} (step_3);
        \draw [draw,->] (step_3) -- node {} (step_4);
        \draw [draw,->] (step_4) -- node {} (step_5);
        \draw [draw,->] (step_5) -- node {} (decision_1);
        \draw [draw,->] (decision_1) -- node [very near start] {Yes} (step_6);
        \draw [draw,->] (step_6) |- node {} (step_7);
        \draw [draw,->] (decision_1) -- node [near start] {No} (step_7);
        \draw [draw,->] (step_7) -- node {} (decision_2);
        \draw [draw,->] (decision_2) -- node [near start] {Yes} (end);
        \draw [draw,->] (decision_2) -| node [pos=0.013] {No} (step_8);
        \draw [draw,->] (step_8) |- node {} (step_2);
    \end{tikzpicture}
    \caption{Sequential Monte Carlo algorithm.}
    \label{fig_smc}
\end{figure}
Figure~\ref{fig_smc} gives an overview of the iterative \gls*{SMC} algorithm. Distributions in \gls*{SMC} are approximated by weighted particles according to
\begin{equation}
    \pi_n(\param) \approx \sum_{i=1}^k w_n^{(i)} \, \delta_{\param_n^{(i)}}(\param),
\end{equation}
wherein $k$ denotes the number of particles and $n$ is the iteration number. Each particle $\delta_{\param_n^{(i)}}$ refers to a Dirac delta distribution at particle location $\param_i$. Distributions $\pi_n$ are not necessarily normalized and, as such, should not be regarded as probability densities. A probability density is approximated with 
\begin{equation}
    \prob_n(\param) \approx \sum_{i=1}^k W_n^{(i)} \, \delta_{\param_n^{(i)}}(\param)
\end{equation}
using the normalized weights
\begin{equation}
    W_n^{(i)} = \frac{w_n^{(i)}}{\sum_{j=1}^k w_n^{(j)}}.
\end{equation}
The algorithm is \textit{initialized} by drawing $k$ particles from the prior distribution with a uniform weight of one
\begin{equation}
    \param_{0}^{(i)} \sim \prob(\param), \,\, w_0^{(i)}=1, \quad \text{with} \quad i\in \mathbb{N}_{>0} , i\leq k.
\end{equation}
It is said that the particles \textit{target} distribution $\pi_0$ (i.e., the prior). By evaluating the likelihood of each particle, the successor distribution can be derived
\begin{equation}
    \pi_{n+1} = \prob\left(y_{\mathrm{obs}}|\model(\param_{n}^{(1,\dots,k)})\right)^{\zeta_{n+1}}.
\end{equation}
The particles are reweighted using importance sampling to target the new distribution $\pi_{n+1}$
\begin{equation}
    w_{n+1}^{(i)} = w_n^{(i)} \cdot \frac{\pi_{n+1}^{(i)}}{\pi_{n}^{(i)}}.
\end{equation}
The exponent $\zeta_n$ refers to the tempering step size and is part of the adaptive tempering reweighting strategy. It is selected based on the \textit{effective sample size} $\mathrm{ESS}$ and the \textit{resampling threshold} $\mathrm{ESS}_{\text{min}}$ according to
\begin{equation}
    \mathrm{ESS}(w_{n+1}^{(1,\dots,k)})\approx \zeta \mathrm{ESS}_{\text{min}}
\end{equation}
with
\begin{equation}
\mathrm{ESS}=\frac{1}{\sum_{i=1}^k\left(w^{(i)}\right)^2}.
\end{equation}
The effective sample size measures how well the weights are distributed between the particles and ensures an appropriate step size for each \gls*{SMC} step. With every iteration, the weights of some particles increase further, and others approach zero. Keeping many particles with (almost) no weights is inefficient, so the particles are \textit{resampled}. Particles with high weights are duplicated, and particles with low weights are dropped. Herein, the systematic resampling strategy from Chopin et al. \cite{chopinIntroductionSequentialMonte2020} is adopted. The particles are resampled based on a multinomial distribution $\mathcal{M}(W_n^{(1,\dots,k)})$ when the effective sampling size $\mathrm{ESS}$ is below the resampling threshold $\mathrm{ESS}_{\text{min}}$. The probability of a particle ${\param^{(i)}}^\prime$ being resampled from particle $\param^{(i)}$ is determined by its relative weight
\begin{equation}
    \prob({\param^{(i)}}^\prime) = \frac{w^{(i)}}{\sum_{j=1}^k w^{(j)}} = W^{(i)}.
\end{equation}
The resampling does not change the distribution targeted by the particles. To avoid a high correlation between particles after resampling (as some particles are duplicates of the same parameter vector), a rejuvenation step is added to move all particles according to
\begin{equation}
    \param_{n+1}^{(i)} = K(\param_{n}^{(i)}),
\end{equation}
using transition kernel $K$ with stationary distribution $\pi_{n+1}$ according to the proposal distribution in an MCMC step. The rejuvenation can be applied multiple times in a row. The process is terminated when the tempering exponent $\gamma_s$ reaches ones, defined as
\begin{equation}
    \gamma_s = \sum_{j=1}^{n+1} \zeta_j.
\end{equation}

\section{Detailed zero-dimensional optimization results\label{sec_detailed_opt_results}}
\begin{figure}[H]
\centering
\includegraphics[width=\linewidth]{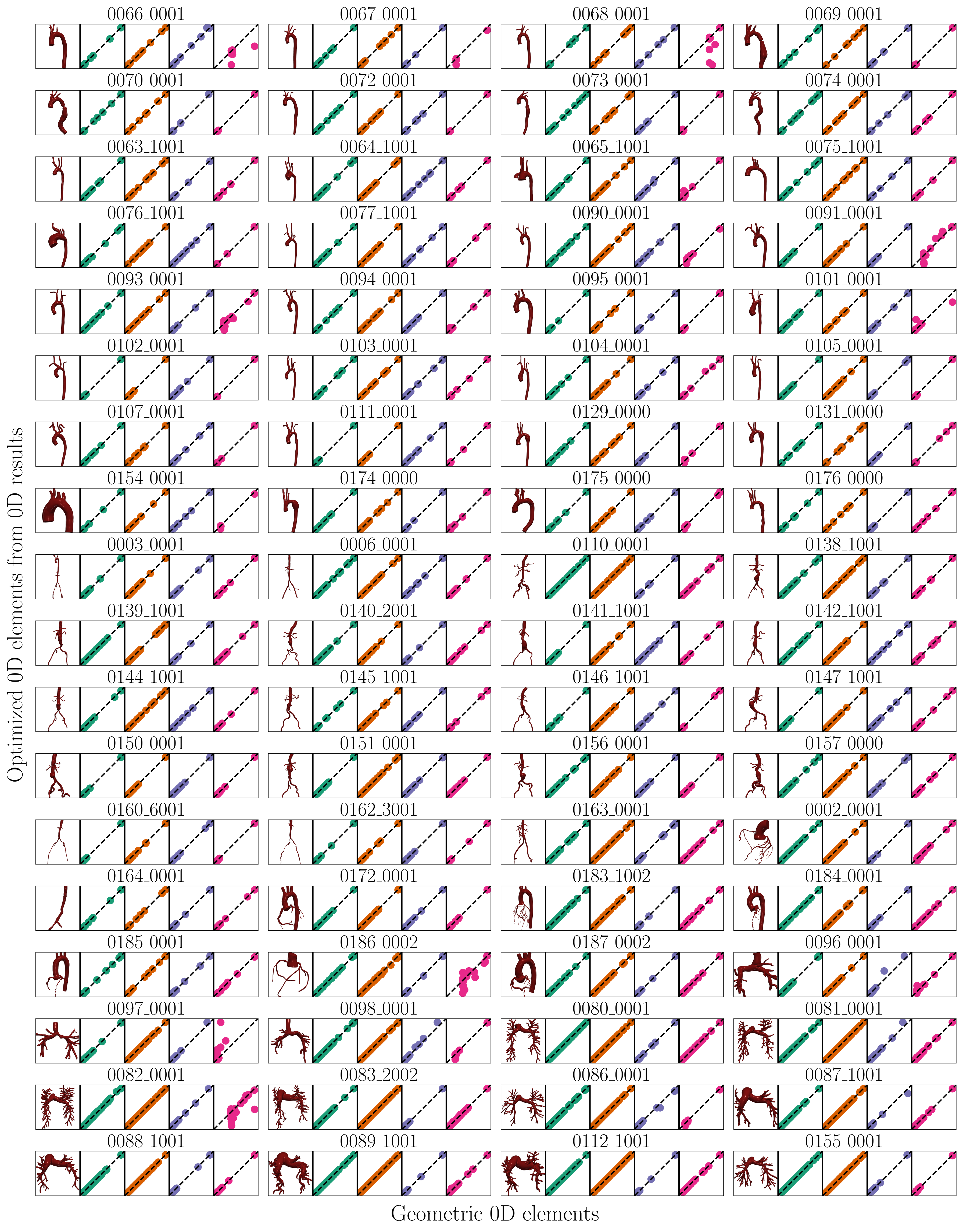}
\caption{Correlation of optimized \gls*{0D} elements from \gls*{0D} ground truth data with \gls*{0D} ground truth for all $N=72$ models (aorta, aortafemoral, coronary, pulmonary). From left to right: $\od{R}$ (green), $\od{L}$ (orange), $\od{C}$ (purple), and $\od{S}$ (pink).\label{fig_optimized_0D}}
\end{figure}

\begin{figure}[H]
\centering
\includegraphics[width=\linewidth]{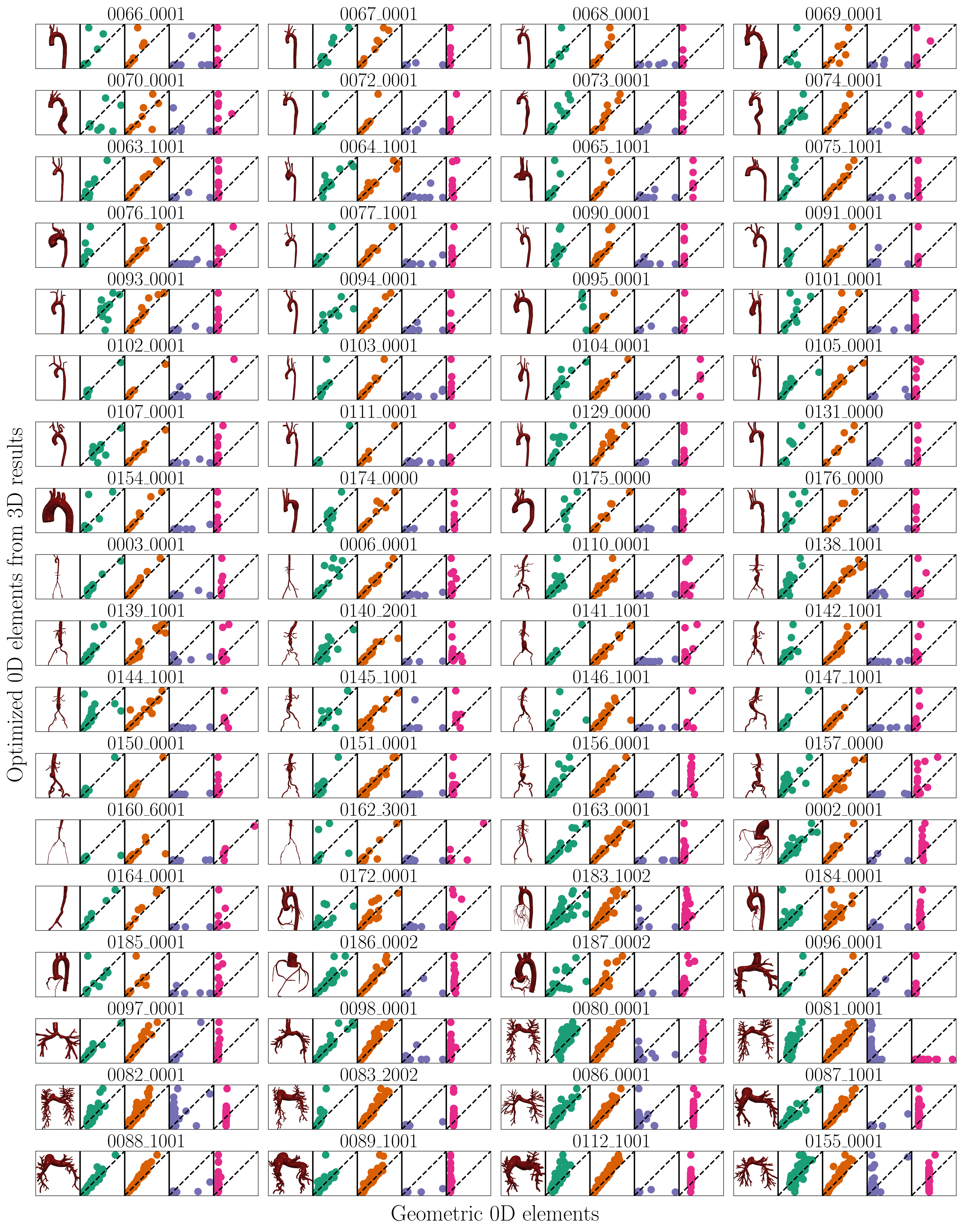}
\caption{Correlation of optimized \gls*{0D} elements from \gls*{3D} data with \gls*{0D} ground truth for all $N=72$ models (aorta, aortafemoral, coronary, pulmonary). From left to right: $\od{R}$ (green), $\od{L}$ (orange), $\od{C}$ (purple), and $\od{S}$ (pink).\label{fig_optimized_3D}}
\end{figure}

The results of the optimization of \gls*{0D} parameters from \gls*{0D} ground truth and \gls*{3D} data are shown in Figures~\ref{fig_optimized_0D} and \ref{fig_optimized_3D}, respectively.

\newpage

\newpage

\newpage
\bibliographystyle{mrp}
\bibliography{references,references_jakob,references_karthik_jonas}

\end{document}